\documentstyle[12pt,psfig]{article}
\begin{document}

\input epsf

\vskip 1truecm
\rightline{Preprint  PUPT-1713, MCGILL-97/24,  DUKE-TH-97-154}
\rightline{ e-Print Archive: hep-ph/9710436}
\vspace{0.2in}
\centerline{\Large Chern-Simons number diffusion with hard thermal loops} 
\vspace{0.3in}

\centerline{\Large Guy D. Moore}

\medskip

\centerline{\it Dept. of Physics, McGill University}
\centerline{\it 3600 University St.}
\centerline{\it Montreal, PQ H3A 2T8 Canada}

\vspace{0.2in}

\centerline{\Large Chaoran Hu and Berndt M{\"u}ller }

\medskip

\centerline{\it Department of Physics, Duke University}
\centerline{\it Durham, North Carolina 27708-0305, USA}

\vspace{0.2in}

\centerline{\bf Abstract}

We construct an extension of the standard Kogut-Susskind lattice model
for classical 3+1 dimensional Yang-Mills theory, in which 
``classical particle'' degrees of freedom are added.  We argue that
this will correctly reproduce the ``hard thermal loop'' effects of hard 
degrees of freedom, while giving a local
implementation which is numerically tractable.  We prove that the
extended system is Hamiltonian and has the same thermodynamics as 
dimensionally reduced hot Yang-Mills theory put on a lattice.  We
present a numerical update algorithm and study the abelian theory to
verify that the classical gauge theory self-energy is correctly
modified.  Then we use the extended system to study the diffusion
constant for Chern-Simons number.  We verify the Arnold-Son-Yaffe
picture that the diffusion constant is inversely proportional to 
hard thermal loop strength.  Our numbers correspond to a diffusion 
constant of $\Gamma = 29 \pm 6 \alpha_{\rm w}^5 T^4$ for 
$m_{\rm D}^2 = 11 g^2 T^2/6$.

\begin{verse} 
\end{verse}

\section{Introduction}
\label{introsec}

One of the most profound and most poorly explained observations of
modern cosmology is that the universe contains macroscopic amounts of
matter, but not of antimatter.  Since baryon number 
is conserved in all observed terrestrial and astrophysical phenomena, the
existence of such a baryon number asymmetry (baryons over antibaryons)
seems peculiar, particularly because it clearly violates
$C$ and $CP$ symmetry.  Further, the abundance of baryons, as compared
to the abundance of photons or the entropy density of the universe, is a
remarkably small number, on order $10^{-10}$; and since the 
entropy of the universe has changed very little since its very hot early
epochs, this small nonzero number is an initial condition to the universe
viewed at least back to the epoch of primordial nucleosynthesis.

Sakharov \cite{Sakharov} made the first attempt to understand what is 
involved in explaining this observation.  If the universe
does not begin with such an asymmetry, then baryon number must be 
violated to generate it; so must $C$ and $CP$.  He also noted that 
{\it if} baryon number is not conserved, then in equilibrium it 
will go to zero, as a consequence of $CPT$ symmetry.
Hence, if the baryon asymmetry of the universe was generated 
dynamically, the universe must in
its early history have gone through some departure from thermal equilibrium,
in which baryon number violation was active and $C$ and $CP$ violating
physics were relevant, followed by a steep suppression of the rate
of baryon number violation before thermal equilibrium resumed (to prevent
the baryon number from being erased again).
And if baryon number was violated at a rate faster than the Hubble 
expansion at any point in the universe's history, then the baryon number
must have been generated dynamically.

While it is known that grand unified theories (GUTs) generically violate 
baryon number, it turns out that baryon number is violated already in the 
standard model, as first shown by t'Hooft \cite{tHooft}.  The key 
observation is that baryon number is coupled through the axial anomaly
to the Chern-Simons number ($N_{\rm CS}$) of the SU(2) weak hypercharge
field, 
\begin{equation}
\frac{dN_{\rm B}}{dt} = N_{\rm F} \frac{dN_{\rm CS}}{dt} \, , 
	\qquad (N_{\rm F} = 3) \, ,
\end{equation}
and that $N_{\rm CS}$ change is thermally activated in the hot
electroweak plasma \cite{Manton,KRS85,ArnoldMcLerran}.  This 
baryon number violation
occurs at much lower temperature ranges, and hence later in the evolution
of the early universe, than GUT mechanisms; 
it could therefore erase any baryon number asymmetry
generated when GUT mechanisms are relevant\footnote{The standard model
violates $B+L$ but not $B-L$.  Extended GUTs such as SO(10) can produce
a $B-L$ excess, which electroweak processes will not touch.  But the
possibility of $B$ production in the standard model certainly motivates
the study of these scenarios.} and generate the observed
abundance during the electroweak phase transition, when the rate abruptly
shuts off.  This latter possibility has stimulated the field of 
electroweak baryogenesis.

One thing we need to know to understand electroweak baryogenesis is 
how quickly baryon number is violated in the standard model,
above the electroweak phase transition.  The violation involves
nonperturbative physics, and there are no known analytic methods
which are reliable above the phase transition temperature 
(in contrast to the situation below it 
\cite{Manton,ArnoldMcLerran}).  Already at the
thermodynamic level the fields responsible for the violation of
baryon number suffer from the ``infrared problem'' of 
thermal field theories with light interacting bosons.

The ``infrared problem'' of the thermodynamics of thermal field theories
with light interacting bosons can be ``solved'' by the dimensional
reduction procedure \cite{oldDR,FKRS1,KLRS}, which reduces the problem
to a three dimensional path integral which can be put on a lattice
\cite{FKRS,Laine,Oapaper} and studied numerically
\cite{KLRSresults,Kripfganz,Teper,MooreTurok,KLRSSU2U1}.  
This ``dimensional reduction''
turns out to be identical to considering the partition function
of the 3+1 dimensional, classical bosonic theory, with certain counterterms
\cite{AmbKras}.  The most hopeful approach to the study of infrared
sensitive, dynamical properties of the plasma is that they also
behave essentially classically \cite{GrigRub}, and there is reason
to believe that they do \cite{MooreTurok,Bodeker}.

There is a problem with the classical approximation to the dynamics,
however.  The ultraviolet modes ($p \sim \pi T$) certainly do not 
behave as classical fields, and they interact with the infrared modes
(which we will assume do behave essentially classically).  The classical
approximation will serve only if the interaction between soft and
hard modes is somehow unimportant to the evolution of the soft modes.
It is known that, at weak coupling, the hard modes only influence the
thermodynamics of the soft modes by shifting the Higgs mass and the
Debye screening mass for the $A_0$ field.\footnote{There
are also perturbatively computable corrections to the couplings of
the theory, which amount to a specification of the renormalization
point, see Farakos et. al. \cite{FKRS1,KLRS}}  However, the
unequal time generalization of Debye screening, the hard thermal
loops, constitute very nontrivial physics, physics which the lattice
implementation of the classical theory gets wrong \cite{Smilga}, both
in the size of the screening effects (which depend inversely on the
lattice spacing) and in the functional details.

Does this problem matter to the rate at which baryon number is violated?
In particular, since there is a fluctuation dissipation relation 
which relates the rate at which a baryon number excess is dissipated
in the plasma to the rate of $N_{\rm CS}$ diffusion per unit 
volume\cite{KhlebShap}, we might ask whether the hard thermal loops
matter to the $N_{\rm CS}$ diffusion constant.  The numerical lattice results
of Ambj{\o}rn and Krasnitz implied that there is a well defined small
lattice spacing limit to the classical, lattice diffusion constant
\cite{AmbKras}, but the definition of lattice $N_{\rm CS}$ used there 
gives bizarre results in the broken phase \cite{TangSmit} due to serious
lattice artifacts \cite{MooreTurok}.  Theoretical arguments suggest
that hard thermal loops should slow down the evolution of infrared 
magnetic fields, for the same reason that infrared magnetic fields in
an abelian plasma become pinned by the conductivity of the plasma
\cite{ArnoldYaffe,HuetSon}; hence one should expect that the $N_{\rm CS}$
diffusion constant vanishes linearly with lattice spacing in classical
simulations.  Recently, numerical techniques based on a topological
definition of $N_{\rm CS}$ \cite{slavepaper} and a ``cooled field'' definition
which removes the worst problems of the old definition \cite{AmbKras2}
have indicated a lattice spacing dependence in the diffusion constant,
though it does not appear to be as large as expected.  Hence, at present
the situation seems confused, and it certainly seems necessary to
find a way to include more faithfully the effects of hard thermal
loops in the study of $N_{\rm CS}$ diffusion, and in other studies of
infrared bosonic field evolution, such as the study of the phase transition
dynamics \cite{MooreTurok} and the generation of quark gluon plasma
in heavy ion collisions \cite{Geiger,Wang,Thoma}.

In this paper we develop, implement, and study 
a proposal by two of us \cite{HuMuller}
for including the hard thermal loop effects into classical, lattice
simulations of Yang-Mills or Yang-Mills Higgs theory.  The implementation
produces a local, Hamiltonian system in which added ``particle''
degrees of freedom convey the hard thermal loop effects.  We explicitly
verify that the enlarged system conserves energy and symplectic measure,
has the same thermodynamics as the quantum theory in the dimensional
reduction approximation, and
produces the right dynamical behavior for infrared fields.  We
examine in detail the retarded propagator of the abelian theory with
particles to show that they correctly produce all of the physics of hard
thermal loops, including Debye screening, plasma oscillations, and
Landau damping.  Then we apply the particle method to determine the
$N_{\rm CS}$ diffusion constant and its dependence on the Debye mass.  The
results verify the arguments of Arnold, Son, and Yaffe (ASY).
In particular, we verify that as the total strength of hard thermal loop
effects varies by a factor of 3, with the lattice spacing and the
physical volume held constant, the diffusion constant for $N_{\rm CS}$
also varies by a factor of 3.

The outline of the paper is as follows.  In Section \ref{HTLsec}
we review what the hard thermal loops are and how they can be understood
in terms of Vlasov equations, ie in terms of the 
influence of a bath of particles on the classical infrared modes.
In section \ref{contintimesec} we explicitly construct a spatial lattice,
continuum time system which can be viewed as an $N$ body simulation of
the Vlasov equations.  We show that this system is energy conserving and
Hamiltonian and argue that, for
small charge and large particle number, it reduces to the (lattice) Vlasov 
equations in the eikonal approximation,
which are known to produce the right hard thermal loops.  In Section
\ref{thermosec} we study its thermodynamics, which are shown to be the
same as dimensionally reduced Yang-Mills theory put on the lattice, 
at a specific Debye mass set by the 
number and charge of the particles.  Section
\ref{algorithmsec} develops a stable, time symmetric update algorithm,
and Section \ref{abeliansec} numerically studies the retarded propagator
of the abelian theory.  Section \ref{NCSsec} uses the evolution 
algorithm, together with the topological definition of $N_{\rm CS}$ 
developed in \cite{slavepaper}, to study the $N_{\rm CS}$ diffusion constant 
and its dependence on lattice spacing and on the size of hard thermal
loop effects.  Section \ref{conclusionsec} concludes.  There are also
four technical appendices.  Appendix A analytically studies how the lattice
nature of the electric fields changes the plasma frequency, in the limit
of many particles of small charge.  Appendix B studies the thermodynamics
of the lattice system when $m_{\rm D} \sim g^2 T/a$.  Appendix C reviews
how to probe the retarded propagator via linear response to an 
external current.  Appendix D discusses the relation between lattice and
continuum time scales.

\section{Hard thermal loops and particles}
\label{HTLsec}

In this section we briefly review an approach to hard thermal loops due to
Kelly et al \cite{KellyMIT94} and discuss how one can generate the influence 
of hard thermal loops by coupling the infrared fields to a heat bath of 
particles.

In the diagrammatic approach to field theory at high temperature it was
shown by Braaten and Pisarski \cite{BraatenPnucl90} that a resummation
procedure is needed in order to take into account consistently all 
contributions at leading order in the coupling constant. Such contributions
are called ``hard thermal loops'' since they arise from one-loop diagrams
with soft external legs and hard internal momenta. The hard thermal loop
resummation was shown to produce gauge invariant results for physical
quantities such as the gluon damping 
rate in a QCD plasma \cite{BraatenPprd90}. 
An effective action for the hard thermal loops was derived by Taylor and
Wong \cite{TaylorWong90} by imposing gauge invariance on the generating
functional. Another approach developed by Blaizot and Iancu 
\cite{BlaizotIprl93} is based on a truncation of the hierarchy of
Schwinger-Dyson equations and the generating functional was obtained 
through a consistent expansion in the coupling constant.

While remarkably insightful, the approaches mentioned above are quite
technical, often involving lengthy calculations, and they 
tend to hide the classical nature of hard thermal loops.  
The hard thermal loops arise from loop diagrams and can be obtained from the
Schwinger-Dyson equations of quantum field theory.  But they are also
UV finite, with loop integrals being exponentially suppressed in the
ultraviolet.  This is because they arise entirely from thermal 
fluctuations.  One usually thinks of such fluctuations, at least in
the ultraviolet, as being well described by classical particles, which
motivated Kelly et al \cite{KellyMIT94} 
to find an alternative, classical formalism for hard thermal loops, by 
considering the linear response of an ensemble of thermal particles to a 
weakly coupled, slowly varying mean field in the framework of classical
transport theory \cite{HeinzElze89}. They start by considering particles
carrying non-abelian SU($N$) charge $q^a, a=1,...,N^2-1$. The Wong equations
\cite{Wong70} are used to describe the proper time evolution of a particle 
with phase space coordinates $(\xi^{\mu}, p^{\mu}, q^a)$:
\begin{eqnarray}
m\frac{d\xi^{\mu}}{d\tau} &=& p^{\mu}~, \label{x}\\
m\frac{dp^{\mu}}{d\tau} &=& gq^aF_a^{\mu\nu}p_{\nu}~, \label{p} \\
m\frac{dq^a}{d\tau} &=& -gf^{abc}p^{\mu}A^b_{\mu}q^c~, \label{q}
\end{eqnarray}
where $g$ is the coupling constant and $F^{\mu\nu}_a$ denotes the strength
of the mean color field $A^{\mu}_a$. Note that the color charge is itself 
subject
to dynamical evolution, a feature which is absent in the abelian case. Both
the dynamical evolution of a particle's spin and the spin coupling
are neglected since spin interactions are down by an order of $g$ in a 
weak mean field describing soft excitations 
($k\sim gT$, $A_{\mu}\stackrel{<}{\sim}T$, $F_{\mu\nu}\stackrel{<}{\sim}gT^2$) 
\cite{Heinz84}.

Consider the classical one particle distribution function $f(x,p,q)$ evolving
in time according to the Boltzmann transport equation,
\begin{equation}
m\frac{df(x,p,q)}{d\tau}=0~, \label{Boltzmann}
\end{equation}
where the collision integral describing hard collisions between particles 
is neglected since soft collisions mediated by the mean 
field dominate. Inserting the Wong equations (\ref{x})-(\ref{q}) into
(\ref{Boltzmann}), one arrives at the following:
\begin{equation}
p^{\mu} \left[ {\partial\over\partial x^{\mu}} - gq^a F_{\mu\nu}^a
{\partial\over\partial p_{\nu}} - gf_{abc} A_{\mu}^bq^c
{\partial\over\partial q^a}\right] f(x,p,q) = 0~. \label{Vlasov1}
\end{equation}
A self-consistent set of non-abelian Vlasov equations for the distribution
function and the mean color field can be obtained by augmenting 
(\ref{Vlasov1}) with the Yang-Mills equations,
\begin{equation}
D_{\mu}F^{\mu\nu} = g \int dpdq p^{\nu}q f(x,p,q) \equiv
j^{\nu}(x)~, \label{Vlasov2}
\end{equation}

The non-abelian Vlasov equations (\ref{Vlasov1}) and (\ref{Vlasov2}) are now
applied to study the soft excitations in a hot, color-neutral plasma with
massless particles. In
the spirit of linear response theory, one expands the distribution function
in powers of $g$,
\begin{equation}
f=f^{(0)}+g f^{(1)}+{\cal O}(g^2)~,
\end{equation}
where $f^{(0)}(p_0)=C n_{\rm {B,F}}(p_0)$ is the equilibrium distribution
in the absence of a net color field. At leading order in $g$, the equation
(\ref{Vlasov1}) reduces to 
\begin{equation}
p^{\mu} \left[ {\partial\over\partial x^{\mu}} - gf_{abc} A_{\mu}^bq^c
{\partial\over\partial q^a}\right] f^{(1)}(x,p,q) 
= p^{\mu}q^a F_{\mu\nu}^a
{\partial\over\partial p_{\nu}}f^{(0)}(p_0)~. \label{Vlasov1_1}
\end{equation}
Similarly, there is a net induced current density in momentum space,
\begin{equation}
j^{\mu a}(x,p)=g^2 \int dq p^{\mu}q^a f^{(1)}(x,p,q)~. \label{current}
\end{equation}

{}From (\ref{Vlasov1_1}) and (\ref{current}), a constraint satisfied by
the color current can be derived. Using the standard field theory
definition of effective action 
$j^{\mu}(x)= -\frac{\delta\Gamma(A)}{\delta A_{\mu}(x)}$,
one then arrives at the hard thermal loop effective action of the
following form:
\begin{equation}
\Gamma_{\rm HTL}=\frac{1}{2} m^2_{\rm D} \left[ \int d^4x
	A_0^a(x)A_0^a(x) -\int\frac{d\Omega}{(2\pi)^4} 
	W(A\cdot v)\right] \, , 
\label{HTL}
\end{equation}
where $v\equiv (1, \mbox{\boldmath $p$}/p_0)$, 
$m_{\rm D}=gT\sqrt{\frac{2N + N_{\rm s} 
+ 2 N_{\rm F}}{6}}$,\footnote{Here, $N_{\rm s}$ is
the number of fundamental scalars, and as in
the preceding text, $N_{\rm F}$ is the number of generations, each containing
four chiral doublets; in QCD 
$2 N_{\rm F}$ would be replaced by $N_{\rm F}$ the
number of flavors, and $N_{\rm s}$ would only appear in supersymmetric
extensions.} 
and the integration 
$\int d\Omega$ is over all directions of the unit vector 
$\mbox{\boldmath $p$}/p_0$. The first term describes Debye screening.
$W(A\cdot v)$ in the second term is a functional. Its explicit form has 
been given by Taylor and Wong \cite{TaylorWong90}, also by Efraty and
Nair \cite{EfratyNair92}. Note that the derivation of $\Gamma_{\rm HTL}$ 
stays completely within the classical transport theory and makes no use 
of the quantum theory. This justifies the statement that 
hard thermal loop effects are classical (in the sense of the classical
particle approximation).  Also note that the form of the
expression inside the brackets in Eq. (\ref{HTL}) does not rely
on the particles obeying any specific statistics; they can
be Fermi, Bose, or even Boltzmann particles.  Nor does it depend on what
group representation they are in.  These only affect how much a species
contributes to the leading coefficient $m_{\rm D}^2$, 
which is given by a sum
over charged species of their individual contributions.  The ratio of a
particle species' contribution to $m_{\rm D}^2$ 
and the mean density of those
particles is $2g^2 C_2(R) \langle E^{-1} \rangle$.  
(For classical, distinguishable
particles in the adjoint representation, $g^2 C_2(R)$ is 
replaced by the mean squared value of the particle charge $q$ in one 
Lie algebra direction.)

It is worth noting that the magnetic field does not play a role
in producing hard thermal loops at leading order. This can be easily seen
from (\ref{Vlasov1_1}), where only the electric field enters the
term on the right. Physically, this is because the magnetic field just
rotates the momentum distribution, and at leading order that
distribution is the rotationally invariant thermal distribution.
The magnetic field only influences existing departures from equilibrium,
which is a subleading effect.  In contrast, the electric field
can polarize the plasma and generate a net current, which in turn 
interacts with the mean field and hence generates the desired hard thermal 
loops. Therefore, in solving the Wong equations, one can leave out the
magnetic term in the Lorentz force in (\ref{p}) 
if the plasma is only slightly driven out of equilibrium. 
However, if one is interested in dynamical 
processes occurring in out-of-equilibrium  plasma, then the force due to
the magnetic field has to be included.

The effective action
(\ref{HTL}) is conceptually simple and formally appealing. It
provides a concise way of summarizing hard thermal loops and
allows one to better understand the influence of the hard thermal modes
on the soft excitations. Nevertheless, it does not prove a ready starting  
point for practical calculations. In particular, its nonlocality 
makes it hard to apply to study nonperturbative physics such as 
Chern-Simons number diffusion, where analytical methods are rare 
and one has to rely on numerical simulations. However, the fact
that the hard thermal loop effective action (\ref{HTL}) can be derived 
from classical transport theory, i.e. the Vlasov equations, implies
that hard thermal loops can be generated by solving the Vlasov equations
numerically.  One could do so by an $N$ body simulation by solving
the coupled system of Wong equations (\ref{x})-(\ref{q}) and 
Yang-Mills equations (\ref{Vlasov2}).
This has the advantage of being local in spacetime and allows a practical,
real time study of both equilibrium and nonequilibrium phenomena. 

Based on these ideas, our goal will be to derive a system of dynamical
equations which, in its low-frequency, long-wavelength limit, reduces
to the same effective field theory as the full thermal gauge theory.
We have no intention of modeling the gauge theory as truly as possible
in all respects; rather, we will be satified to reproduce its infrared
behavior correctly. For example, it will be irrelevant for us whether
the system of classical particles carrying gauge charge has detailed
resemblance to the hard thermal modes of the gauge field, if only its
back-reaction produces the nonlocal HTL effective action (\ref{HTL}). The
system of classical dynamical equations to be derived in the following
sections will have three parameters: the classical thermal length
scale $(g^2T)^{-1}$, the gauge charge $(gQ)^2$ of the particles, and
the particle density $\langle n\rangle$. It will be sufficient for our
purpose if we tune these parameters so that they produce the correct
values for the parameters in the HTL action.

\section{The lattice system}
\label{contintimesec}

To reproduce the hard thermal loop effects, our idea is then to
numerically implement ``particles'' obeying Wong's equations and moving in
the background of classical, lattice fields.  Eventually we want a 
discrete time update algorithm, but an important step is to 
construct a continuous time, lattice system with the right properties.
The general philosophy of adding
particles with adjoint (Lie algebra) charge $q$ of fixed
magnitude, kinetic momentum $\vec{p}$, and continuous position coordinate
$\xi$ satisfying $\dot{\xi}_i = p_i/|p|$, has been presented
in \cite{HuMuller}; here we will
specify the complete implementation.  Our requirements for this system
are
\begin{itemize}
\item 	There must be conservation of energy,
\item 	The evolution should preserve the phase space (symplectic) measure,
\item	The system should respect cubic, (lattice) translation, $C$, $P$, 
	and $T$ symmetry,
\item	Gauss' law should be identically preserved,
\item	The small lattice spacing $a$ limit (or smooth field, large time
	limit if one thinks in lattice units) must recover the
	Yang-Mills field equations supplemented with Wong's equations,
\item	The thermodynamics of the infrared classical fields
	must be given by the path integral of quantum Yang-Mills
	theory in the dimensional reduction approximation (or its
	lattice discretization).
\end{itemize}
Note that the first two conditions ensure that the system is
Hamiltonian, and hence that the thermodynamics are well defined.  We
would also like a thermalization algorithm for the system.

We emphasize that the particles are a device to reproduce the hard
thermal loops and should not be taken literally as reproducing all the
behaviors of the hard modes.  In particular it is not a problem that
they are distinguishable, and that their number is conserved.  We expect
them to satisfy Boltzmann statistics, rather than Bose or Fermi-Dirac
statistics, but this is also not important to reproducing the functional
form of the hard thermal loops, as discussed in the last section; we
need only make sure that the number density and charge of the particles
yields the desired value for $m_{\rm D}^2$.  However, any hope that
the method can be enhanced to account for physics beyond hard thermal
loops is clearly remote.  Hence we will not be able to say
anything about effects which are subleading in $g$.

\begin{figure}[t]
\centerline{\psfig{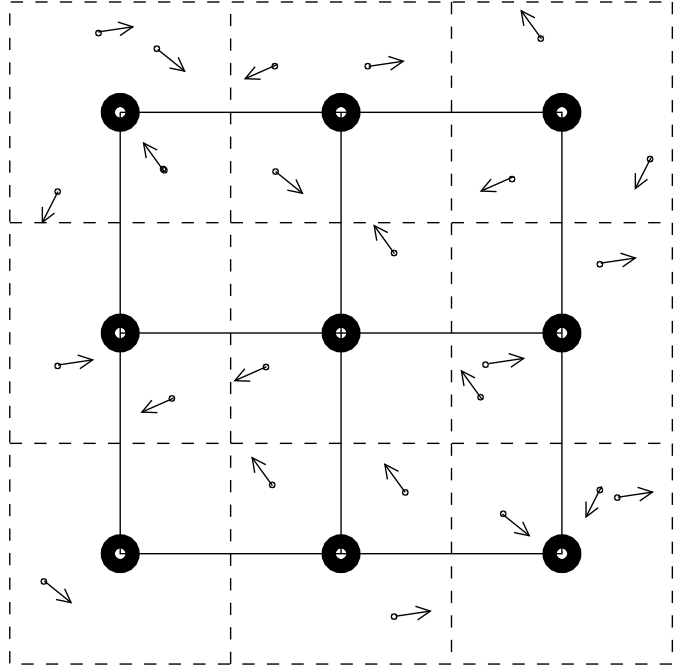}}
\caption{\label{illustration}
An illustration of the degrees of freedom of the proposed system.
Lattice sites are large dots, and the solid lines joining them are
links.  Classical fields take values at sites (Higgs
fields) or on links (connections $U$ and
electric fields $E$).  Particles (the small dots)
take on real valued coordinates and momenta (illustrated with arrows).  
A particle's charge affects the classical fields as if it resided at the
nearest lattice site.  The dotted lines (really planes, extending out of
the page) are barriers between the region
nearest one site and that nearest another, that is, faces of the dual
lattice.  When a particle crosses a
barrier, the charge is parallel transported to the new box, 
the $E$ field on the
link orthogonal to the barrier receives a kick, and the particle
momentum orthogonal to the barrier is changed to conserve energy.}
\end{figure}

Let us list the degrees of freedom of the proposed system 
(see Fig.~\ref{illustration} for illustration).  We consider
a 3-torus of spatial extent $N^3$ (in lattice units, which will be used
throughout except when it is convenient to write the lattice spacing
explicitly).  ``The lattice'' will refer to the integer lattice on this
space, ie all points with all three coordinates an integer.  A link
$x,i$ will refer to the line between the lattice site $x$ and the site
$x + \hat{i}$ (henceforward $x+i$) and on each link there will be a
parallel transporter $U_i(x) \in$ SU(2) and an electric field
$E_i(x) \in$ LSU(2), the Lie algebra of SU(2).  For each index value
$\alpha \in \{1,..,N_{\rm p} \}$ there is a 
particle with coordinate $\xi_\alpha$
defined on the torus, momentum $p_\alpha \in \Re^3$, and
charge $q_\alpha \in$ LSU(2) satisfying $q^2_\alpha = Q^2$.  By 
definition $E_i(x)$ will be the left acting covariant time derivative
of $U_i(x)$, $D_0 U_i(x) = E_i^a(x) i \tau^a U_i(x)$, and
the momentum $p_{\alpha}$ will tell the direction the particle moves in,
$\dot{\xi}_{\alpha,i} = p_{\alpha,i} / |p|_{\alpha}$ (note the 
$\xi_i$ are defined mod $N$).  It remains to define update rules
for $E$, $p$, and $q$.

First, what is the meaning of $q$?  It should be a charge which the
classical fields $U,E$ ``see,'' but such a charge should reside at a lattice 
point.  We take $q_\alpha$ to ``live'' at the lattice point $x_i$
closest to $\xi_\alpha$, ie $|x_i - \xi_{\alpha,i}| \leq 0.5$.  It will
gauge transform as an adjoint object at that site;
and the charge observed by the classical fields will be
\begin{equation}
\rho(x) = \sum_\alpha q_\alpha \times \left\{ \begin{array}{cl}
1 & \xi \; {\rm at} \; x \\ 0 & \xi \; {\rm not \; at} \; x \\
\end{array} \right. \, .
\label{defrho}
\end{equation}
This is the quantity which enters Gauss' law,
\begin{equation}
\rho(x) = \sum_i E_i(x) - U^{\dagger}_i(x-i) E_i(x-i) U_i(x-i) 
	\equiv D_{\rm L} \cdot E(x) \, .
\label{Gausslaw}
\end{equation}

The current $j^a_i$, an adjoint vector field, 
should be defined on the links and
is specified by the requirement that it form a conserved current,
\begin{equation}
\dot{\rho} = \sum_i U^{\dagger}_i(x-i) j_i(x-i) U_i(x-i) - j_i(x) \, .
\end{equation}
The easiest way to see how this fixes the current is to think of the charge 
$q_\alpha$ of particle $\alpha$ as living at site $x$ until the particle
moves to be closest to another site, say $x+i$; then the charge must 
abruptly slide along the link connecting $x$ and $x+i$, and there
will be a ($\delta$ function in time) current on that link,
equal to $j_i(x) = q_\alpha \delta(t-t_{\rm cross})$.
If we take the group indicies of
$j$ to live at the basepoint of the link $x$, then we should use the value
of $q$ which departs from site $x$; 
the value if we take $j$ to live at
the endpoint of the link $x+i$ is the adjoint parallel transport of $j$
with group indicies at the basepoint, and hence the charge $q$ of the 
particle when it arrives 
at site $x+i$ must be the parallel transport of the value at $x$,
\begin{equation}
q({\rm at} \;x+i) = U^{\dagger}_i(x) q({\rm at} \; x) U_i(x) \, .  
\label{qtransform}
\end{equation}
If the particle is moving the other direction, 
from $x+i$ to $x$, the sign of the current is reversed, but 
(\ref{qtransform}) still holds.  This 
gives the update rule for $q$ and is the same as was proposed in
\cite{HuMuller}.

In what follows we will call such events ``boundary crossings'' because
they correspond to a particle crossing the boundary which demarks the
volume closest to one lattice site.  The boundary 
is a face of the dual lattice,
dual to the link in which the current flows.

We expect that the electric field update rule should be
\begin{eqnarray}
\label{Eupdate_contin}
\frac{dE_i(x)}{dt} & = & - \frac{\partial H_{\rm KS}}{\partial U_i(x)} 
	- j_i(x) \, , \\
H_{\rm KS} & \equiv & \sum_{x,i} \frac{E_i^2(x)}{2} + 
	\sum_{\Box} \left( 1 - \frac{1}{2} {\rm Tr} U_{\Box} \right) \, ,
\end{eqnarray}
where $H_{\rm KS}$ is the Kogut-Susskind Hamiltonian \cite{Kogut} which is
standard in the real time field literature
\cite{Ambjornetal}, and
where $\partial/\partial U_i(x)$ means change with respect to left acting
derivatives of $U_i(x)$.  The appearance of $j$ here is the ``meaning''
of $j$ and just means that $E_i(x)$ changes abruptly by $-q_\alpha$ when
particle $\alpha$ moves from nearest site $x$ to nearest site $x+i$, and
by $q_\alpha$ if it goes the other way.

\begin{figure}
\centerline{\psfig{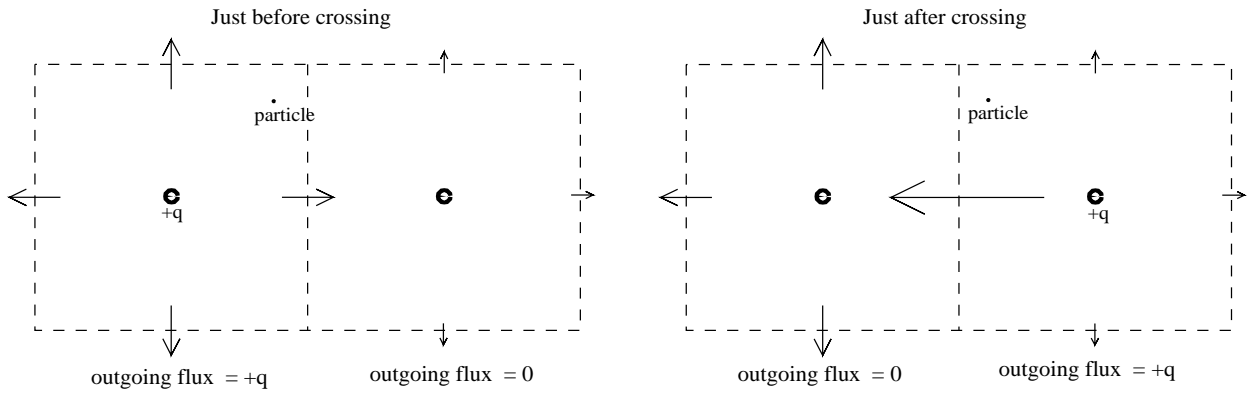}}
\caption{ \label{boundarycrossfig}
Boundary crossing, before and after.  Before, the flux of $E$ fields
(displayed as arrows) out
of the lefthand box must equal $q$, and the flux from the righthand box 
must equal 0.  After the crossing, the righthand box has flux $q$ and
the lefthand box has zero.  This demands a change of the $E$ field 
connecting them, but no instantaneous change to
the $E$ fields which leave these two boxes, since that would change the
flux going into other boxes.}
\end{figure}

Now let us check whether this definition of $\dot{E}$ will preserve
Gauss' law, that is, whether
\begin{equation}
\dot{\rho}(x) = \sum_i \dot{E}_i(x) - U^{\dagger}_i(x-i) \dot{E}_i(x-i) 
	U_i(x-i) \, .
\label{Gausschange}
\end{equation}
(The contribution from the time derivative of $U$ cancels here because
it is generated by, and hence commutes with, $E$, 
and cancels between $U$ and $U^{\dagger}$.)

First, we know that
\begin{equation}
\sum_i \frac{\partial H_{\rm KS}}{\partial U_i(x)} - 
	U_i^{\dagger}(x-i) \frac{\partial H_{\rm KS}}{\partial U_i(x-i)}
	U_i(x-i) = 0 \, ,
\end{equation}
which is why Gauss' law is preserved for the classical lattice system
without particles.
And the contribution of $j$ to $\dot{E}$ cancels $\dot{\rho}$ 
in (\ref{Gausschange}) precisely
because $\rho$ and $j$ form a conserved current.  So Gauss' law is indeed
conserved.  Alternately, the abrupt change to $\rho(x)$ and $\rho(x+i)$
when a particle passes from being closest to $x$ to being closest to $x+i$
stipulates the change to $E_i(x)$ so that Gauss' law will still hold
at both sites.  

Physically, what this update of $E$ means is that 
when a particle leaves one box, the
charge within that box abruptly drops, and the flux of $E$ field out of
the box must abruptly drop too.  Meanwhile, the neighboring box, which
received the particle, must have the flux of $E$ field out of it
abruptly rise.  The way to do this is to abruptly change the $E$ field
going from the first box to the second, by an amount equal to minus the
charge, see Fig. \ref{boundarycrossfig}.  
We could also do it by changing the $E$ fields along the other
links leading out of these boxes, but this would spoil Gauss' law at
the sites at the other ends of those links 
and is therefore forbidden.  Those links will change, in time, via
the Hamiltonian evolution of the classical Yang-Mills system.

We should mention that in the abelian theory, one
can construct more elaborate Gauss law preserving
ways to change $E$, which boil down to
moving some of the charge along indirect paths between $x$ and $x+i$.
But these are actually not allowed in the nonabelian theory, because the
charge transforms nontrivially.  To satisfy Gauss' law, the particle
charge $q$ must also be updated by splitting it up 
and parallel transporting it along those
indirect paths; but when they are added together in the final box, they
will not all ``point in the same Lie algebra direction'', so the
magnitude of the sum
will have changed.  Similar problems arise if one tries to
define the charge of a particle as being spread over nearby points,
rather than residing entirely at the nearest point.
Deriving a local, Gauss' law preserving, $C$, $P$, and $T$ symmetric 
alternative to our update proposal is highly nontrivial in the
nonabelian context.

Before going further we should make a comment about the normalization of
fields which we have adopted.  Since $U$ is a parallel transporter it
has been most convenient to use the normalization of electric fields in
which the Hamiltonian contains a $1/g^2 \times E^2$, that is our
electric field is $g a^2 E_{\rm cont} / 2$.  The $1/g^2$ in front of
the Hamiltonian will be absorbed into the temperature, 
$\beta_{\rm L} = 4 / (g^2 a T)$, which is customary in the 3-D lattice
literature \cite{AmbKras}.  Also the particle charge $Q^2$ will convert
into ``usual'' continuum charge as $Q^2 = g^4 q^2 / 4$.  
It will be convenient
to make the particle momentum appear in the Hamiltonian as $\beta_{\rm L} H =
\beta_{\rm L} ( H_{\rm KS} + \sum_\alpha |p|_{\alpha} )$, so that it can be
directly compared with electric field energies.  It is then
related to physical units by $p_{\rm latt} = p_{\rm cont}
\times g^2 a / 4$.  Though these normalizations seem strange, they scale
out all dimensionful quantities and make the numerical theory have the
least awkward inter-relations.

It remains to define the update of $p$.  This will be almost
uniquely specified by the requirement that the system be Hamiltonian.
Naively one would expect the influence of the electric field on $p$
to be
\begin{equation}
\dot{p}_i = q^a E_i^a \, ,
\end{equation}
with $E_i$ chosen to be $E$ on the nearest $i$ type link; but this is
wrong, as the system energy then has a time derivative of 
$p_i q E_i / |p|$, which cannot be removed by a corresponding change in 
$E$ because that would spoil Gauss' law.  The energy conserving update
\begin{equation}
\frac{dp_i}{dt} = q^a E_i^a - p_i \frac{ p_j q^a E_j^a}{p^2}
\end{equation}
does not preserve the phase space measure and is also disallowed; in fact,
except at the instant when a particle crosses a boundary, measure and energy
conservation restrict the allowed changes to $p$ to a rotation about some
adjoint charged vector $B$, which must be odd under $C$, $P$, and $T$;
\begin{equation}
\dot{p}_i = f(p^2) \epsilon_{ijk} p_j B_k^a q^a \, .
\label{rotate}
\end{equation}
So a rotation about the magnetic field is allowed, but not required,
to preserve energy and the phase space measure.

However, when a particle crosses the face of the dual lattice separating
two points $x$ and $x+i$ and induces a current $j_i(x)$, it changes
the energy in the electric field $E_i(x)$ by
\begin{equation}
\Delta({\rm energy}) =  \Delta ( E^2/2 ) = 
	-E_i(x,t_{\rm cross} - 0) \cdot q + q^2/2
	= -E_i(x,t_{\rm cross} + 0) \cdot q - q^2/2 \, ,
\label{energyshift}
\end{equation}
and $p$ must be changed to balance this energy.
If only $p_i$ changes, from $p_{i,{\rm init}}$ to 
$p_{i,{\rm fin}}$, then energy conservation is
\begin{equation}
\sqrt{ p_{i,{\rm fin}}^2 + p_\perp^2} + \Delta ( E^2/2 )
	= \sqrt{ p_{i,{\rm init}}^2 + p_\perp^2} \, ,
\end{equation}
which is solved by
\begin{equation}
p_{i,{\rm fin}} = {\rm sign}(p_{i,{\rm init}}) \sqrt{ ( |p|_{\rm init} - 
	\Delta ( E^2/2 ) )^2 - p_\perp^2 }
\label{outmomentum}
\end{equation}
if both $|p|_{\rm init} - \Delta ( E^2/2 ) \geq 0$ and the argument of
the square root is $\geq 0$; otherwise there is no solution, the 
crossing is energetically forbidden.  If the crossing is energetically
forbidden, we should set 
$p_{i,{\rm fin}} = - p_{i,{\rm init}}$, no crossing occurs, no current is
generated, and the particle turns.  Otherwise, $p_{i,{\rm fin}}$ is taken
from (\ref{outmomentum}) above, the current flows, and the particle 
crosses the boundary.  The case where the crossing is from $x+i$ to $x$
follows from this case and parity symmetry.

This proposed update for $p$ conserves energy.  What 
about symplectic measure?  The measure is
clearly preserved at all times that no particle crosses a boundary,
since the evolution of $E,U$ is Hamiltonian, $\dot{\xi}_\alpha$ depends on
$p_\alpha$ but not $\xi_\alpha$, and $q$ and $p$ do not change (except
for allowed rotations of $p$).  When a particle reflects, the change
$p_i \rightarrow - p_i$ also preserves the measure.  The only nontrivial
case is when a particle crosses a boundary.  The rotation of $q_\alpha$
preserves the measure on LSU(2) and the change $E \rightarrow E - q$
preserves the measure for $E$ since the change is independent of $E$,
which is defined on a vector space.  We need only check if the particle
phase space measure $\prod_i dp_i d\xi_i$ is preserved, that is, whether
\begin{equation}
{\rm Det} \left[ \begin{array}{cc} 
	\partial p_{i,{\rm fin}} / \partial p_{j,{\rm init}} &
	\partial p_{i,{\rm fin}} / \partial \xi_{j,{\rm init}} \\
	\partial \xi_{i,{\rm fin}} / \partial p_{j,{\rm init}} &
	\partial \xi_{i,{\rm fin}} / \partial \xi_{j,{\rm init}} \\
	\end{array} \right]
\end{equation}
equals 1.
Since the change we are discussing is instantaneous, 
$\partial \xi_{i,{\rm fin}}
/ \partial p_{j,{\rm init}} = 0$ and the determinant reduces to blocks,
\begin{equation}
{\rm Det} \left[ \begin{array}{cc} 
	\partial p_{i,{\rm fin}} / \partial p_{j,{\rm init}} &
	\partial p_{i,{\rm fin}} / \partial \xi_{j,{\rm init}} \\
	\partial \xi_{i,{\rm fin}} / \partial p_{j,{\rm init}} &
	\partial \xi_{i,{\rm fin}} / \partial \xi_{j,{\rm init}} \\
	\end{array} \right] = 
{\rm Det} \left[ \frac{ \partial \xi_{i,{\rm fin}}}
	{\partial \xi_{j,{\rm init}}} \right] 
	{\rm Det} \left[ \frac{ \partial p_{i,{\rm fin}}}
	{\partial p_{j,{\rm init}}} \right] \, ,
\end{equation}
and we need to show that these determinants are inverses.

Without loss of generality we consider the case in which
$p_1$ changes, with $p_1 > 0$.  We handle the dependence of
$\vec{\xi}_{\rm fin}$ on $\vec{\xi}_{\rm init}$ first.  
Changing $\xi_{2,{\rm init}}$ just
shifts the place on the wall where the particle crosses, but $\vec{p}$
is the same at all times; so only $\xi_{2,{\rm fin}}$ is changed, by
the same amount as the initial change.  The same holds for $\xi_3$.
However, changing $\xi_{1,{\rm
init}}$ by $d\xi_1$ changes the arrival time at the wall by $- d\xi_1
/ v_{1,{\rm init}}$.  The particle propagates at $\vec{v}_{\rm fin}$
rather than $\vec{v}_{\rm init}$ for $d\xi_1 / v_{1,{\rm init}}$ longer than
without the change, leading to a change in the final positions equal
to $(\vec{v}_{fin}-\vec{v}_{init}) d\xi_1/ v_{1,{\rm init}}$, plus the 
change of $d\xi_1$ in $\xi_{1,{\rm fin}}$.  Hence,
\begin{equation}
{\rm Det} \left[ \frac{ \partial \xi_{i,{\rm fin}}}
	{\partial \xi_{j,{\rm init}}} \right] =
	{\rm Det} \left[ \begin{array}{ccc}
	v_{1,{\rm fin}} / v_{1,{\rm init}} & 0 & 0 \\
	(v_{2,{\rm fin}} - v_{2,{\rm init}})/v_{1,{\rm init}} & 1 & 0 \\
	(v_{3,{\rm fin}} - v_{3,{\rm init}})/v_{1,{\rm init}} & 0 & 1 \\
	\end{array} \right] = \frac{v_{1,{\rm fin}}}{v_{1,{\rm init}}}
	\, .
\end{equation}

Now for the dependence of $\vec{p}_{\rm fin}$ on $\vec{p}_{\rm init}$.
Since $p_{2,{\rm init}} = 
p_{2,{\rm fin}}$ and similarly for $p_3$, only the change in the 
final value of $p_{1,{\rm fin}}$ need be computed.  Taking the
appropriate derivatives of (\ref{outmomentum}) gives
\begin{eqnarray}
{\rm Det} \left[ \frac{ \partial p_{i,{\rm fin}}}{\partial p_{j,{\rm init}}}
	\right] & = &
	{\rm Det} \left[ \begin{array}{ccc}
	\frac{p_{1,{\rm init}}}{p_{1,{\rm fin}}} 
	\left( \frac{ |p|_{\rm init} 
	- \Delta(E^2/2)}{|p|_{\rm init}} \right) & 
	- \frac{\Delta(E^2/2) p_2}{p_{1,{\rm fin}} |p|_{\rm init}} &
	- \frac{\Delta(E^2/2) p_3}{p_{1,{\rm fin}} |p|_{\rm init}} \\
	0 & 1 & 0 \\ 0 & 0 & 1 \\ \end{array} \right] \\
	& = & \frac{p_{1,{\rm init}}}{p_{1,{\rm fin}}} 
	\left( \frac{ |p|_{\rm init} 
	- \Delta(E^2/2)}{|p|_{\rm init}} \right) \, . \nonumber
\end{eqnarray}
Now observe that $|p|_{\rm init} - \Delta(E^2/2) = |p|_{\rm fin}$
and that $p_1/|p| = v_1$; so the two determinants are inverses and the
measure is indeed conserved.  The cases where $p_2$ or $p_3$ change,
or where the particle moves in the opposite direction, follow from cubic
and parity symmetry.  

If we had allowed $p$ in the
orthogonal directions to change, it generically would not preserve
the measure, so the choice of update is almost unique.\footnote{A rotation
of $p_\perp$ violates parity unless its magnitude depends on $B$ or some
other $P$ odd field; and an energy conserving change in the magnitudes
of $p_1$ and $p_\perp$ will have nonunity Jacobian.}

The first four conditions for a valid update rule have specified
the update uniquely except for the freedom to perform rotations of the
form shown in (\ref{rotate}).  It would seem natural to define the
magnetic field $B^a_i$ at a point in terms of the 3 nearest 
plaquettes\footnote{One should be careful here to make the choice of
plaquettes in a way which is preserved under parity and the cubic point
group.} and to rotate momenta according to
\begin{equation}
\dot{p}_i = \epsilon_{ijk} \frac{p_j}{|p|} B_k^a q^a \, .
\end{equation}
As we discussed, the magnetic term in Wong's equations plays no role in
reproducing the right hard thermal loops for a plasma close to equilibrium.  
Thus we are free to omit this rotation when dealing with quasi-equilibrium
processes.

Next let us verify that the continuum limit of our update rules
give Wong's equations.  This is known for the classical field equations 
without particles evolving under the Kogut-Susskind Hamiltonian, so
we need only check the terms involving particles.  The update of the 
particle charge is explicitly the adjoint parallel transport along
its trajectory, as it should be.  The current in the
$i$ direction from a particle occurs in jolts of magnitude $q$ and
frequency $1/v_i$, so the time averaged current is $q v_i$ as it should be.
The impulse on a moving particle is of magnitude $\Delta p_i = 
p_{i,{\rm fin}} - p_{i,{\rm init}}$; expanding 
(\ref{energyshift}) and (\ref{outmomentum}) to leading order in
$q$ gives $\Delta p_i = q \cdot E_i / v_i$, and such impulses also
occur with frequency $1/v_i$, so the time averaged force on the particle
is $q \cdot E_i$, also as it should be.  We will explore the corrections
to the plasma frequency,
due to the discrete nature of the current and the ``kicks,'' 
in Appendix A.  Wong's equations are recovered for
the motion of a particle through slowly varying fields, if $q$ is
sufficiently small.

We still must check whether the thermodynamics of this system are 
correct; we will do this in the next chapter.  But first, let us
review and summarize the update rule.  The fields evolve under the
Kogut-Susskind Hamiltonian and the particles move freely, except at 
such exact instants when a particle crosses a boundary (a face of 
the dual lattice), i.e.~when it goes from being nearest one point to 
nearest another.  The charge is then parallel transported by the link
operator dual to the face, and the electric field on that link
is abruptly changed
by $-q$ (or $+q$ if the motion is in the $-$ direction).  The particle
momentum parallel to the link is changed to cancel the energy change
of the electric field.  If that is energetically impossible, the particle
momentum parallel to the link is flipped and no crossing, or change to
$q$, $E$, occurs.  This update is Hamiltonian with total energy
\begin{equation}
{\rm Energy} = \sum_{x,i} \frac{E_i^2(x)}{2} + 
	\sum_{\Box} \left( 1 - \frac{1}{2} {\rm Tr} U_{\Box} \right)
	+ \sum_{\alpha} |p|_{\alpha} \, ,
\end{equation}
and preserves Gauss' law identically.  It is also manifestly gauge, $C$,
$P$, $T$, and cubic invariant.

\section{Thermodynamics of the lattice system}
\label{thermosec}

In the last section we have proposed a continuum time system and its
update rule and have shown that it is Hamiltonian, ie that the update
rule preserves energy and phase space measure.  It therefore has well
defined thermodynamics, which we now explore.  The canonical partition
function is
\begin{eqnarray}
Z & = & \int \prod_{\alpha} d^3\xi_{\alpha} d^3p_{\alpha} dq_\alpha
	\prod_{i,x} dE_i(x) dU_i(x) \prod_x \delta ( - \rho(x) +
	D_{\rm L} \cdot E (x)) e^{- \beta_{\rm L} H} \, , \\
H & = & \sum_{x,i} \frac{E_i^2(x)}{2} + 
	\sum_{\Box} \left( 1 - \frac{1}{2} {\rm Tr} U_{\Box} \right)
	+ \sum_{\alpha} |p|_{\alpha} \, .
\end{eqnarray}
The appropriate measures, ie the Haar measure for $U$ and the Lebesgue
measure on LSU(2) restricted to $q^2 = Q^2$ for $q$, are implied.  The
$\delta$ function enforces Gauss' law at each point and the meanings
of $\rho$ and $D_{\rm L} \cdot E$ are as previously defined.
The quantity $\beta_{\rm L}$ here combines all the dimensionful parameters
of the lattice system, $\beta_{\rm L} = 4 / g^2 a T$; when it is large we
are on a fine lattice, or at weak coupling, which is equivalent in the
classical theory.

Note first that the kinetic momenta will obey a Boltzmann distribution
and are independent of all other degrees of freedom; the partition
function factorizes and if we are interested in the thermodynamics of the 
IR classical gauge fields then we may integrate out $p$.  Note also that
$\xi$ and $q$ do not appear in the Hamiltonian; their only influence is
in determining $\rho$ appearing in Gauss' law.

It is most convenient to enforce Gauss' law with an adjoint valued 
Lagrange multiplier $A_0$ \cite{AmbKras}, 
\begin{equation}
\prod_x (D_{\rm L} \cdot E(x) - \rho(x) ) = \int \prod_x dA_0(x) 
 	\exp\left[ i \beta_{\rm L} \sum_x A_0^a(x) 
 	(D_{\rm L} \cdot E^a(x) - \rho^a(x) ) \right] \, ,
\end{equation}
which makes the electric fields quadratic; the $E$ integral may now
be performed, generating a kinetic term for $A_0$.  The partition 
function reduces to
\begin{eqnarray}
Z & = & \prod_{i,x} dU_i(x) dA_0(x) e^{- \beta_{\rm L} H_{UA}} I(A_0) \, , \\
H_{UA} & = & \sum_{\Box} \left( 1 - \frac{1}{2} {\rm Tr} U_{\Box} \right) +
	\frac{1}{2} \sum_x (D_{\rm L} A_0)^2(x) \, , \\
I(A_0) & = & \int \prod_{\alpha} d^3\xi_{\alpha} dq_\alpha e^{-i 
	\beta_{\rm L} \sum_x A_0(x) \rho(x)} \, ,
\end{eqnarray}
where the meaning of $(D_{\rm L} A_0)^2$ should be clear.  The particles 
enter in the last term, which only depends on $A_0$.  

Now let us compute $I(A_0)$.  Since $\rho$ is the sum of $\rho_\alpha$ from
each particle, $\exp{-i \beta_{\rm L} A_0 \rho}$ is the product over $\alpha$
of $\exp{-i \beta_{\rm L} A_0 \rho_\alpha}$; the integral 
factorizes into an integral over each particle,
\begin{equation}
I(A_0) = \prod_\alpha \left[ \int dq_\alpha \int d^3\xi_\alpha
	\exp \left( - i \beta_{\rm L} \sum_x A_0^a(x) q_\alpha^a 
	\times \left\{ \begin{array}{cl} 1 & \xi \; {\rm at} \; x \\
	0 & \xi \; {\rm not \; at} \; x \\ \end{array} \right.
	\right) \right] \, .
\end{equation}
The integral over $\xi$, normalized so $\int d\xi = 1$,
gives a sum over sites of a term where the 
particle is at that site,
\begin{equation}
I(A_0) = \prod_\alpha \left[ \frac{1}{N^3} \sum_x \int dq_\alpha
	\exp( - i \beta_{\rm L} A_0^a(x) q^a_\alpha ) \right] \, .
\end{equation}
Now using
\begin{equation}
\int dq \exp( - i \beta_{\rm L} A_0^a(x) q^a ) = 
	\frac{ \sin \left( \beta_{\rm L}
	Q \sqrt{A_0^2(x)} \right)}{\beta_{\rm L} Q \sqrt{A_0^2(x)}} \, ,
\label{justinSU2}
\end{equation}
(normalizing so that $\int dq = 1$) and performing the product,
we arrive at
\begin{equation}
I(A_0) = \left[\frac{1}{N^3} \sum_x \frac{ \sin \left( \beta_{\rm L}
	Q \sqrt{A_0^2(x)} \right)}
	{\beta_{\rm L} Q \sqrt{A_0^2(x)}} \right]^{N_{\rm p}} \, .
\label{Ia0}
\end{equation}

The above expression is exact but not very insightful as written.  It
is best to take a thermodynamic limit and to expand the expression in
the sum.  Denote by $\langle {\rm ARG} \rangle$ the mean value of
the argument of the sum in (\ref{Ia0}).  Rescale $I(A_0)$ by 
$\langle {\rm ARG} \rangle^{-N_{\rm p}}$, which just changes the
normalization of the partition function, and write it as
\begin{equation}
I(A_0) = \left[  1 + \left( -1 + 
	\frac{1}{N^3 \langle {\rm ARG} \rangle}
	\sum_x \frac{ \sin \left( \beta_{\rm L}
	Q \sqrt{A_0^2(x)} \right)}
	{\beta_{\rm L} Q \sqrt{A_0^2(x)}} \right) \right]^{N_{\rm p}} \, .
\end{equation}
As $N^3 \rightarrow \infty$ and $N_{\rm p} \rightarrow
\infty$ with $N_{\rm p} / N^3 \equiv \langle n \rangle$ fixed, the
term in round parenthesis 
vanishes as $N^{-3/2}$, and we may use the approximation
\begin{equation}
\left( 1 + \frac{x}{N} \right)^{N}
	= \exp(x) \exp(-x^2/2N) \times ( 1 + O( x^3/N^2)) \, .
\end{equation}
The $\exp(-x^2/2N)$ term in the identity means that there 
is a very weak nonlocal interaction term between fluctuations from the
mean value of 
$A_0^2$ at pairs of points, trying to force the global average of $A_0^2$
towards its equilibrium value.  It is probably safe to ignore this term, 
and higher corrections
strictly vanish in the large $N$ limit.  Neglecting the nonlocal
term, we get
\begin{equation}
I(A_0) \simeq \exp \left( \frac{\langle n \rangle}
	{\langle {\rm ARG} \rangle} \sum_x \frac{ \sin \left( \beta_{\rm L}
	Q \sqrt{A_0^2(x)} \right)}{\beta_{\rm L} Q 
	\sqrt{A_0^2(x)}} \right) \, .
\end{equation}
Since we are typically interested in a system where $N^3 \sim 10^4$ and
$N_{\rm p} > 10^5$, the thermodynamic limit is justified.

Now, expanding the term in the sum,
\begin{equation}
\frac{ \sin \left( \beta_{\rm L}
	Q \sqrt{A_0^2(x)} \right)}{\beta_{\rm L} Q \sqrt{A_0^2(x)}}
	= 1 - \frac{\beta_{\rm L}^2 Q^2 A_0^2(x)}{6} + 
	\frac{\beta_{\rm L}^4 Q^4 (A_0^2)^2(x)}{120} - \ldots \, ,
\end{equation}
we see that $I(A_0)$ contains a Debye screening term and quartic and
higher self-interaction terms for $A_0$; $I(A_0)$ becomes
$\exp ( - \beta_{\rm L} V(A_0^2))$, where the potential $V$ is
\begin{eqnarray}
V(A_0^2) & = & \frac{m_{\rm D}^2}{2} \sum_x A_0^2 + \frac{\lambda_{\rm A}}{4}
	\sum_x (A_0^2)^2 + \ldots \, , \\
m_{\rm D}^2 & = & \frac{ \langle n \rangle \beta_{\rm L} Q^2}
	{3 \langle {\rm ARG} \rangle} \, , \label{valueofmd}\\
\lambda_{\rm A} & = & - \frac{ \langle n \rangle \beta_{\rm L}^3 Q^4}
	{30 \langle {\rm ARG} \rangle} \, , \,\ldots \, .
\end{eqnarray}
Finally, 
\begin{equation}
\langle {\rm ARG} \rangle = 1 - \frac{ \beta_{\rm L}^2 Q^2}
	{6} \langle A_0^2 \rangle + O(\beta_{\rm L}^4 Q^4 A_0^4) \, .
\end{equation}
$\langle A_0^2 \rangle$ is UV dominated and well approximated by its
perturbative value.  At
lowest order, $\langle A_0^2 \rangle = 3 \Sigma / 4 \pi \beta_{\rm L}$,
with $\Sigma \simeq 3.1759$ \cite{Laine}.  In practice
it is necessary to make $Q \sim \beta_{\rm L}^{-1}$, see below, 
in which case, for $\beta_{\rm L} \sim 10$,
$\langle {\rm ARG} \rangle$ typically differs from 1 by $\sim 1 \%$.
As we mentioned in Section \ref{HTLsec}, the ratio 
$m_{\rm D}^2 / \langle n \rangle$ should equal 
$\langle 2 q_1^2 \rangle \langle
E^{-1} \rangle = (Q^2 / 3) (2 \beta_L / 2)$ for classical particles in the
continuum.  Discretization has shifted this by a small $O(Q^2)$ correction.

In a group other than SU(2), the calculation begins to go differently at
Eq. (\ref{justinSU2}), and the final Debye mass is different.  For
instance, in U(1) gauge theory, the integral there gives
$\cos(\beta_{\rm L} Q A_0)$, and $m_{\rm D}^2$ 
and $\lambda_{\rm A}$ are 3 and 5 times larger, respectively.

We have now shown that the thermodynamics are the 
same as lattice regulated Yang-Mills
theory in the dimensional reduction approximation, at a specific
value of Debye mass, except for the added higher order interaction
terms for the $A_0$ field.  A minimal requirement for the thermodynamics
to be all right is that the $A_0$ field is not strongly coupled,
$\lambda_{\rm A} \leq 1$.  The lattice $\lambda_{\rm A}$ corresponds to the
physical $4 \lambda_{\rm A} / g^2$, so this requirement is that the coupling
of the $A_0$ sector is weaker than the coupling of the 
gauge sector, which should
be sufficient since the $A_0$ field is quite massive.  This requirement
is roughly
\begin{equation}
Q^4 \langle n \rangle < \frac{30}{\beta_{\rm L}^3} \, .
\end{equation}
In fact it will turn out that dynamic considerations require that
$Q$ be on order or smaller than $\beta_{\rm L}^{-1}$, in which case,
unless $\langle n \rangle$ is very large, $\lambda_{\rm A}$ will be very
small, as it should be in the dimensionally reduced Hamiltonian.

We should also require that the Debye screening mass from particles
be larger than that from hard lattice modes (as otherwise the hard
thermal loops are dominated by wrong lattice mode contributions, rather
than right particle contributions), in which case
\begin{equation}
\frac{12 \Sigma}{\pi \beta_{\rm L}^2} < \langle n \rangle Q^2 \, .
\label{mDconstraint}
\end{equation}
However, it is not necessary on thermodynamic grounds to make $m_{\rm D}^2$
small in lattice units.  When $m_{\rm D}^2$ is large in physical units, ie
$m_{\rm D}^2 \gg \beta_{\rm L}^{-2}$ in lattice units, 
then the influence of the 
$A_0$ field on the vector fields is perturbative.  Making $m_{\rm D}^2$ on
order the lattice spacing complicates the problem of integrating out
the $A_0$ field to find its influence on the infrared physics, but the
one loop integration can still be performed; we treat this problem
in Appendix B.  However, it will turn out on dynamical grounds that one
should not make the Debye mass too large, as the plasma frequency then
takes on lattice artifact corrections, discussed in Appendix A.

\section{Discrete time update algorithm}
\label{algorithmsec}

In this section we give a detailed description of a stable and accurate
discrete time update algorithm to solve the equations
of motion for the particles and lattice fields. 
Probably most readers can safely skip this section; nothing in it is
essential for understanding the rest of the paper, although it certainly
is important that a stable and well behaved algorithm exists for
implementing the system described in the last two sections.

The numerical algorithm we will construct 
is essentially a leapfrog, with modifications to meet the
needs of the specific problem under investigation. In particular we 
make sure the algorithm is time centered, and the particle update part is
exactly energy conserving.  The stepsize errors in the algorithm should
be $O((\Delta t)^2)$ where $\Delta t$ is the time step in lattice units.

The variables to be updated are: $(\xi_{\alpha}, p_{\alpha}, q_{\alpha})$
and $(E_{x,i}, U_{x,i})$. 
If the connections $U$ were fixed and the particles did not rotate due to
magnetic fields, then we could perform the update (of $E$, $p$, $q$, $\xi$)
exactly, as follows.
Starting with $E$, $p$, $\xi$, $q$ at time $t$, we project where $\xi$ 
will be at time $t + \Delta t$,
\begin{equation}
\xi_{{\rm proj},\alpha,i}(t + \Delta t) = \xi_{\alpha,i}(t) + 
	\Delta t \frac{ p_{\alpha,i}}{|p|_{\alpha}} \, .
\end{equation}
For all $\alpha$ where the projected location is in the same box (dual 
lattice cell) as the initial location, ie no boundary is crossed, then
$\xi(t+\Delta t)$ equals the projection, and we update $\xi$.  For
other points, we draw the straight line path between $\xi(t)$ and 
$\xi_{\rm proj}$ and find the first boundary it crosses.  $\xi$ can
be updated to this point, $\xi_{\rm bound}$, which it reaches 
at time $t_{\rm bound} = t + | \xi_{\rm bound} - \xi(t) |$.
At this point we must
be careful, because more than one particle may interact with the same
electric field, and the updates will depend on the order of
interaction.  So we order all particles which cross a boundary
according to crossing time; then we solve the crossing conditions in
that order, modifying $E$, $q$, and $p$ as discussed in Section
\ref{contintimesec}.  For each particle, after these 
variables are updated, the
particle's position at time $t + \Delta t$ is again projected,
starting at $\xi_{\rm bound}$ and using the new $p$.  If it crosses
another boundary before time $t + \Delta t$, we again update it to the
boundary, compute the time of arrival, and insert it at the place
appropriate for the new crossing time in the collection of particles
to be updated; otherwise we update it to its projected position at
time $t+\Delta t$.  When the last crossing has been dealt with, then all
$E$, $p$, $q$, and $\xi$ have been updated to time $t + \Delta t$.
This algorithm is exact except for roundoff errors.

One comment is in order about this update.  Ordering the particles by
time of crossing takes $O(N_{\rm p} \ln N_{\rm p})$ steps, 
so the algorithm does
not quite scale linearly with volume.  In practice, though, the 
ordering takes
$C \Delta t N_{\rm p} \ln N_{\rm p}$ computations 
with $C$ a fairly small number,
and this part of the algorithm takes less time than formally $O(N_{\rm p})$
or $O(N^3)$ parts.

Now we must incorporate the above idea into the leapfrog update of the
$E$, $U$ fields.  The leapfrog in the absence of particles is
\begin{eqnarray}
U_i(x,t + \Delta t/2) & = & \exp( i \Delta t \tau^a E^a_i(x,t)) U_i(x , 
	t - \Delta t/2) \, , \label{Uupdate_Disc} \\
E_i^a(x,t+\Delta t) & = & E_i^a(x,t) - \Delta t \frac{\partial H_{\rm KS}
	( U(t + \Delta t/2) )}{\partial U_i(x,t + \Delta t/2)} \, ,
	\label{Eupdate_Disc}
\end{eqnarray}
where the meaning of $\partial H_{\rm KS} / \partial U$ is defined between
(\ref{Eupdate_contin}) and (\ref{Gausschange}).  
If one needed to define $E_i^a(x,t +
\Delta t/2)$ one could do so by applying only half of the update,
(\ref{Eupdate_Disc}).  

We combine this leapfrog and the update of $\xi$ etc. discussed above,
as follows.
\begin{enumerate}

\item
Start with $E(t)$, $p(t)$, $q(t)$, $\xi(t)$, and $U(t - \Delta t/2)$.

\item
Determine $U(t+\Delta t/2)$ according to (\ref{Uupdate_Disc}).

\item
Apply half the $E$, $U$ leapfrog update of $E$; namely, set
\begin{equation}
E_i^a(x,t+0) = E_i^a(x,t) - \frac{\Delta t}{2} \frac{\partial H_{\rm KS}
	( U(t + \Delta t/2) )}{\partial U_i(x,t + \Delta t/2)} \, ,
\end{equation}
Also update $p(t)$ to $p(t+0)$ by rotating about $B$ for time 
$\Delta t/2$; first set
\begin{equation}
p_{{\rm temp},\alpha,i} = p_{\alpha,i} + \frac{\Delta t}{2}
	\epsilon_{ijk} \frac{p_{\alpha,j}}{|p|_{\alpha}} B^a_k(\xi)
	q^a \, .
\end{equation}
(Here $B_k(\xi)$ means the $i,j$ plaquette which is closest to the
point $\xi$ and with indicies living at the point where the indicies
of $q$ reside.)  Then rescale $p$ to its original magnitude,
\begin{equation}
p_{\alpha,i}(t+0) = p_{{\rm temp},\alpha,i} \frac{|p|_{\alpha}(t)}
	{|p|_{{\rm temp},\alpha}} \, .
\end{equation}

\item
Update $E$, $q$, $\xi$, and $p$ from time $t+0$ to time $t+ \Delta t -
0$ using the ``fixed connection, no $B$ field'' algorithm presented
above, and $U = U(t + \Delta t/2)$.

\item
Apply the other half of the $E$, $U$ leapfrog,
\begin{equation}
E_i^a(x,t+\Delta t) = E_i^a(x,t+\Delta t - 0) - 
	\frac{\Delta t}{2} \frac{\partial H_{\rm KS}
	( U(t + \Delta t/2) )}{\partial U_i(x,t + \Delta t/2)} \, ,
\end{equation}
and rotate the momenta as in step 3.  Now, return to $1.$, but with the
value of $t$ incremented by $\Delta t$.

\end{enumerate}

Applying these in order is one leapfrog update.  Note that the update
is time symmetric and exactly Gauss constraint preserving at each
step, and that step $4.$ exactly conserves energy.  The overall
conservation of energy is exactly as good as in the Kogut-Susskind
leapfrog algorithm, that is, energy fluctuates by a small $O( (\Delta
t)^2 N^{3/2})$ amount (which for a $20^3$ 
grid at $\Delta t = 0.05$ and $30$ particles per site is less
than a part in $10^5$ of the total energy) 
and the central value of the energy is absolutely stable.

As we have discussed, the effect of magnetic fields on particles 
does not contribute to the hard thermal loops for a plasma 
close to equilibrium.  
So that part of steps $3.$ and $5.$ can be left out when studying 
quasi-equilibrium processes, which in practice saves at least $1/3$
of the update time.  

Finally we present a canonical ensemble thermalization algorithm for
this system.  It is a straightforward generalization of the
``constrained molecular dynamics'' algorithm of \cite{Moore1}.
Beginning from an arbitrary choice of $U(\Delta t/2)$, $\xi(0)$, 
and $q(0)$, we choose $p(0+0)$ from the Boltzmann distribution at some
inverse temperature $\beta_{\rm L}$ and $E(0+0)$ from the Gaussian
distribution at the same temperature, modulo the Gauss constraints.
As in \cite{Moore1} this is done by choosing $E$ without regard to the
constraints and then orthogonally projecting to the constraint
surface, which correctly thermalizes the transverse components of $E$
and correctly enforces the constraints on the longitudinal
components.  The algorithm is identical to that in \cite{Moore1}
except that the particle contribution to Gauss' law must be added.
This chooses $p$, $E$ with thermal weight from the fixed $U$, $\xi$,
$q$ subspace of phase space.  Then we evolve the system under the
Hamiltonian evolution, using the algorithm presented above, for some
length of time, at the end of which we again draw $E$ and $p$ from
the thermal ensemble; we repeat until the (athermal) original information in
the $U$, $\xi$, and $q$ has been destroyed and measurables attain
values which do not change in the mean under Hamiltonian evolution or
further thermalization.  The spirit of this algorithm is that of a
molecular dynamics Monte-Carlo.  Note that it is important to choose
$E$ and $p$ at some time between $t+0$ and $t+\Delta t - 0$ and not at
$t$ or $t + \Delta t$ because the half update from $t-0$ to $t$
changes $E$ from being Gaussian and uncorrelated with $\partial H_{\rm KS}
/ \partial U$ to being correlated with $\partial H_{\rm KS}
/ \partial U$.  (Similarly in \cite{Moore1} it was necessary to choose
$E$ defined at the same time as $U$ and to perform a half update
before beginning the leapfrog.)

We should also note that there is no obstacle to applying a Langevin
type thermalization algorithm based on the one developed in
\cite{Krasnitz}, and in particular that it is trivial to couple
Langevin noise to the particle momenta.  This might be important if one
wanted to simulate thermalization of these modes through interactions
with some other degrees of freedom, for instance strong scattering of
fermions in the electroweak model.

\section{Some results for the abelian theory}
\label{abeliansec}

Before diving into the study of $N_{\rm CS}$ diffusion we should check
that the particle method is producing the right physics of HTL's.  In
this section we will first discuss how small $Q^2$ must be for the
system to give good behavior, and then we will study the abelian
theory at suitably small $Q^2$ to see if the hard contributions to the
retarded self-energy are correct.

\begin{figure}[t]
\centerline{\psfig{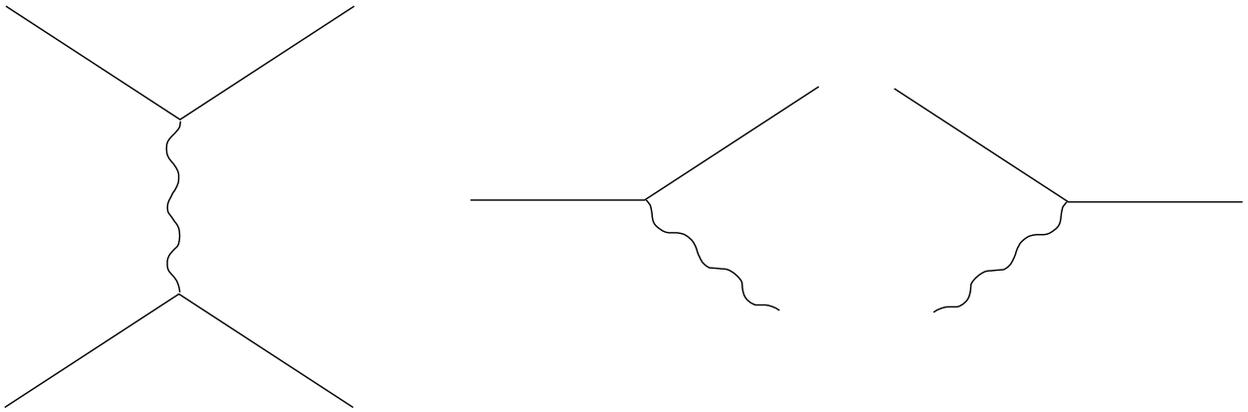}}
\caption{\label{scattfig}
Diagram which naively dominates particle scattering, left, and Cerenkov 
diagrams which are really dominant, right.}
\end{figure}

We saw in Section \ref{thermosec} that a necessary condition for the
theory to have a weakly coupled $A_0$ sector is 
\begin{equation}
Q^4 \langle n \rangle < \frac{30}{\beta_{\rm L}^3} \, .
\end{equation}
In fact, dynamical considerations demand that $Q^2$ be still smaller.
To get the right hard thermal loop effects in the dynamics, we need a
particle to travel a distance longer than the magnetic length scale
$1/g^2T$ (or $\beta_{\rm L}$ in lattice units) before its momentum
is randomized by interactions with the plasma.  Normally we would
expect the dominant randomizing process to be Coulomb scattering from
other particles as depicted on the left-hand side of Figure
\ref{scattfig}; but in fact the dominant processes are absorption or
emission of hard classical field excitations.  These processes are not
kinematically forbidden because the lattice dispersion relations ``turn
over'' at high momentum, so hard excitations move slower than the
speed of light, allowing particles to Cerenkov radiate or absorb.  From
the point of view of the lattice modes, this is saying that the
ultraviolet lattice modes are
Landau damped due to the particles,
which is possible because their dispersion relation has
$ \omega / k < 1$.

A ``worst case'' estimate is that the typical electric field which a
particle interacts with is uncorrelated with the charge $q$ of the
particle, so the momentum $p$ falls by $q^2 / 2 = Q^2 / 2$ in addition
to receiving a kick uncorrelated with its initial value.  The average
particle (averaging over directions of motion) crosses walls with
frequency 3/2 in lattice units; so the decorrelation rate for $p$
under the above approximation is
\begin{equation}
{\rm rate} \sim \frac{3 Q^2}{4 \langle p \rangle} = \frac{ Q^2
	\beta_{\rm L}}{4} \, ,
\end{equation}
which must be $\ll \beta_{\rm L}^{-1}$; so $Q^2 \ll 4 / \beta_{\rm L}^2$.  In
practice the above estimate for randomization is close to the real
behaivor; for $\beta_{\rm L} = 10$ and $Q = 0.08$, the decoherence time is
about $80$ in lattice units.  In our work we typically demand $Q^2 \leq
\beta_{\rm L}^{-2}$, which forces the number of particles to be quite large,
$\langle n \rangle \sim 30$, in order to satisfy
(\ref{mDconstraint}).  This makes the update of particles the
dominant numerical cost, but it pushes us closer to the Vlasov
equation limit.

Note also that the processes mentioned above make the evolution of 
the UV classical lattice modes damped and noisy, 
with a damping strength proportional to $Q^2 \langle n \rangle$.
The consequences for the infrared magnetic sector deserve investigation.

The above behavior is disturbing enough to encourage us to check that
the behavior of the infrared degrees of freedom is correct.  This is
most easily done for the abelian theory, because there the electric
field and the current are gauge invariant and one can easily probe the
system with external currents and study the response.  It would also
be possible to do this in the nonabelian theory in a specific gauge,
but complications from pure classical gauge theory interactions
complicate interpreting results; the abelian theory provides a nice,
clean environment to study the dynamics of the particle technique. 

We will perform two numerical tests on the abelian theory.  First, we verify 
that the lattice field has the correct dispersion in the infrared.
Second, we probe the retarded propagator by studying the linear response 
of the abelian plasma to an external current.

\begin{figure}[t]
\centerline{ \epsfxsize=12cm \epsfbox{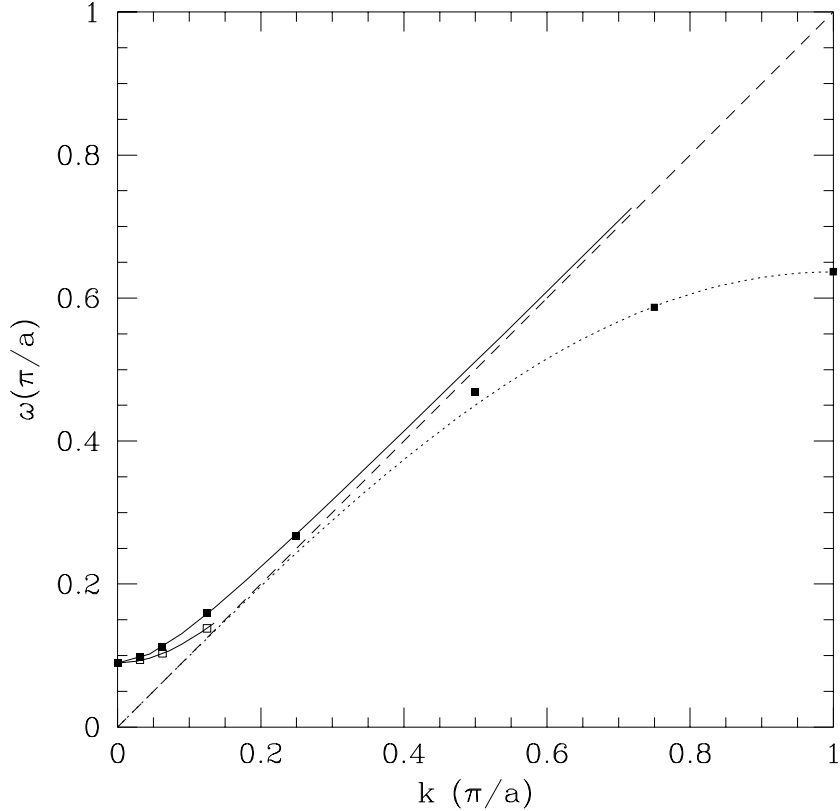} }
\caption{\label{dispersion}
Dispersion relation: free field in the continuum (dashed line: 
$\omega = k$); free field on the lattice (dotted curve: 
$\omega a = 2\sin{(k a/2)}$); plasma modes: transverse (theory:
the upper solid curve; data: solid rectangles) and longitudinal
(theory: the lower solid curve; data: open rectangles).}
\end{figure}

To measure the plasma dispersion relation, we pick initial conditions
so the (spatial) Fourier mode of wave vector $k$ has a large electric
field.  This sets up a plasma oscillation with this wave number.  We
evolve the fields and measure the electric field in the same Fourier 
mode at regular time intervals.  Then we Fourier transform this time
series and pick out the frequency with the most power; this is the
plasma frequency.  We have used this procedure to measure the dispersion
for both transverse and longitudinal plasma
modes. The results are summarized in Figure \ref{dispersion}, where we
compare them to the lattice, free field dispersion relation and the
continuum dispersion relations, with and without hard thermal loops.
These data are for $Q = 0.0159$, $\beta_{\rm L} = 18.86$, and
$\langle n \rangle = 50$, corresponding to a plasma frequency of
$\omega_{\rm p} = 0.282$.  (Recall that in the abelian theory
$m_{\rm D}^2$ and $\omega_{\rm p}^2$ are 3 times larger in 
terms of $Q^2 \langle n \rangle$ than in SU(2).)
The numercal results agree with theory 
remarkably well in the infrared. As $k$ gets bigger, the data points 
deviate from the continuum theory curve and bend down to match the 
lattice dispersion relation. 
This is expected because hard plasma modes are much 
less influenced by hard thermal loops and behave like free lattice modes.

We have also computed the plasma frequency at $k=0$ analytically, 
in Appendix \ref{AppendixA}.
There we conclude that the correction due to lattice artifacts
should be negligible for
$\omega_{\rm p}^2 \ll 1$ in lattice units. This is supported by our 
numerical results.

To probe the retarded (transverse) photon propagator, we drive the system 
with an external current of the following form:
\begin{equation}
j_i= \hat{j}_i j_0 \sin{(\omega t)} \sin{(k x)} \, ,
\end{equation}
where the amplitude $j_0$ is small so that linear response theory applies.
We study response of the plasma and measure the space-time average of
$j\cdot E$, which, according to Appendix \ref{AppendixC}, can be written
in terms of the transverse photon polarization function:
\begin{equation}
\langle j\cdot E\rangle = \frac{j_0^2}{4}\cdot
\frac{\omega \Pi_{\rm i}(\omega,k)}
{\left[\omega^2-k^2-\Pi_{\rm r}(\omega,k)\right]^2 + \Pi^2_{\rm i}
(\omega,k)} \, ,
\label{je}
\end{equation}
where $\Pi_{\rm r}(\omega,k)$, $\Pi_{\rm i}(\omega,k)$ 
are the real and imaginary
parts of the transverse polarization function, respectively.  We also
measure the out of phase response of $E$ and the case of longitudinal
excitation; the specific expressions are in Appendix \ref{AppendixC}.

\begin{figure}
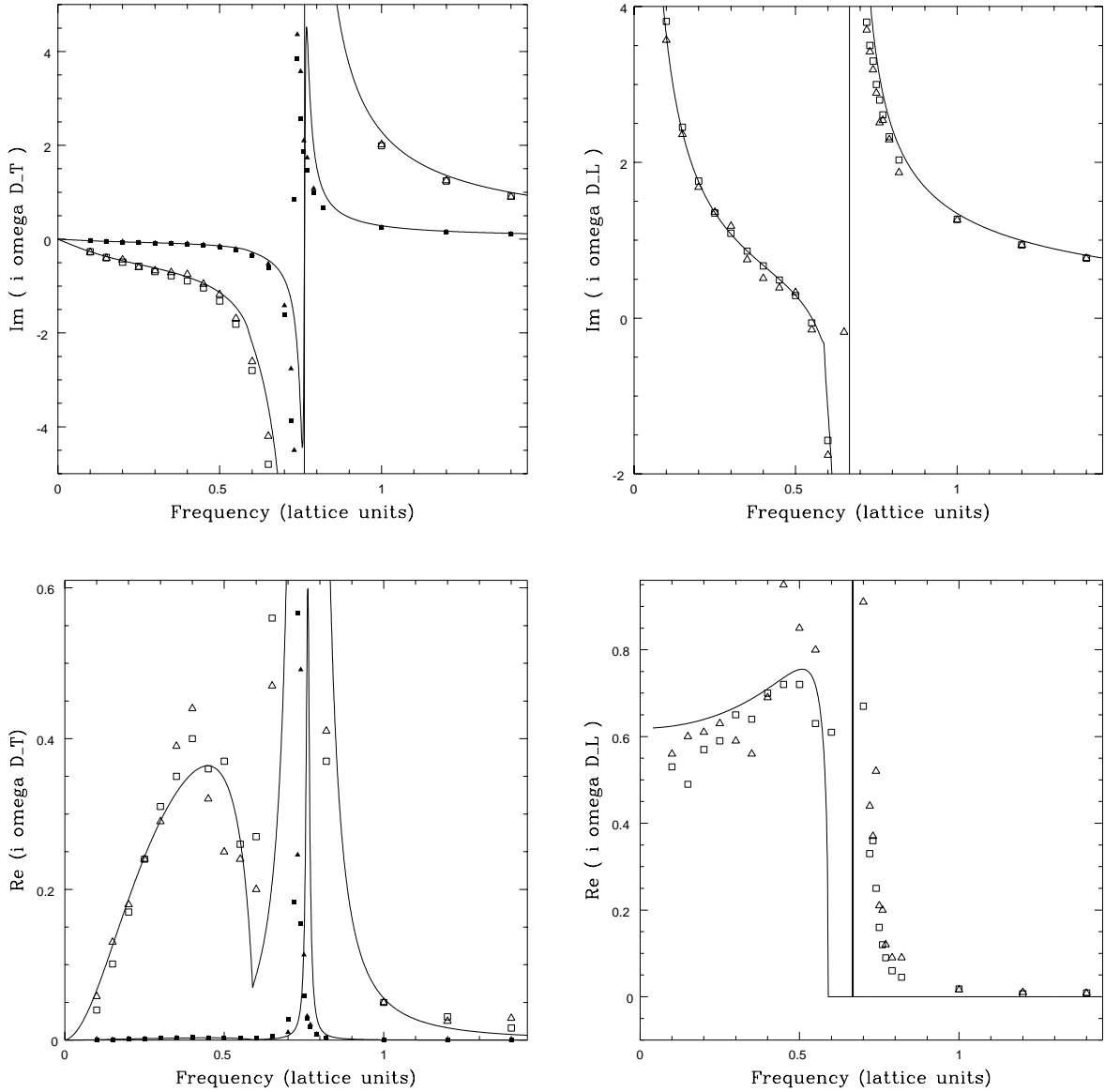

\centerline{\mbox{\psfig{file=t3out.epsi,width=2.9in}}
\hspace{0.2in}
\mbox{\psfig{file=l3out.epsi,width=2.9in}}}
\vspace{0.3in}
\centerline{\mbox{\psfig{file=t3in.epsi,width=2.9in}}
\hspace{0.2in}
\mbox{\psfig{file=l3in.epsi,width=2.9in}}}
\caption{\label{dat_vs_th}
Retarded Propagator--actually $i \omega D^{\rm R}$.  
Above left, the out of phase
(nondissipative) part of the transverse propagator; lower left, the in
phase (dissipative) part of the transverse propagator.  The right
figures are the same for the longitudinal propagator.  In each case, the
solid lines are theory, the squares are the $k=(3,0,0) \times \pi/16$
data, and the triangles are the $k=(2,1,1) \times \pi/16$ data.  In the
transverse theory figures, the curves with smaller, filled data
points are rescalings of the data and theory, so the resonance will fit in
the plot; the rescalings are by 8 and 120 for the out of phase and in
phase responses respectively.  All numbers are in lattice units.
}
\end{figure}

We want both to compare to theory and to test the rotational invariance
of the particle induced hard thermal loops.  To do so, we choose two $k$
vectors which are of the same length but are inequivalent under the cubic 
point group, and we study the complex propagator for each, at
frequencies above and below the plasma resonance.  We choose the
vectors $k = (3,0,0) \times \pi/16$ and $k = (2,2,1) \times \pi / 16$ on
a $32^3$ lattice, with $\beta_{\rm L} = 20$, $Q^2 = 0.0005$, and
$\langle n \rangle = 60$ (so $\omega_{\rm p} = 0.447$).  We excite the
plasma at each wave vector and numerous frequencies in turn and
integrate the resulting response for long enough to get reasonably clean
results.  

We present the results, plotted against the theory, in Figure
\ref{dat_vs_th}.  A few comments are in order.  In the strict hard thermal
loop approximation there would be no imaginary part to the self-energy
above the light cone, but the appearance of a small imaginary part,
which will arise at $O(Q^4 \beta_{\rm L}^2 \langle n \rangle)$ since we
are in the abelian theory, qualitatively
changes the response near the resonance by giving the resonance some
width.  For the transverse data, we have added a small phenomenological
imaginary part to the self-energy chosen to make the theory give a
resonance of about the same sharpness as the data.  The match between
data and theory is very good except that the location of the resonance
is shifted a little towards lower $\omega$ in the data.  This is
expected; the particle response weakens with frequency because the
particles cannot respond to a field faster than they encounter it, which
is how often they cross a boundary, as we examine quantitatively in 
Appendix A.  Also note that the data for $k \propto (2,2,1)$ (triangles)
have their resonance very slightly after the data for $k \propto (3,0,0)$,
which is expected from the $O(k^4)$ lattice corrections to the
dispersion relations.  However, besides this, they are
indistinguishable, which is a good check that the particle contribution
to the self-energy is rotationally invariant.

We have not added a small imaginary part to the self-energy above the
light cone in the theory lines in the plots for the longitudinal
propagator, but the data clearly show
that one is present.  The agreement between data and theory is
generally good for both longitudinal and transverse propagators, and the
data for the two values of $k$ disagree by about the same amount as
the jitter in the data caused by statistical error.  The comparison to
the theory without hard thermal loops is stark; for instance, the
in phase response would then be zero in all cases, 
and the out of phase, longitudinal
response would have no resonance but would behave simply as $1 /
\omega$.  

We conclude that the particles add hard thermal loops which, for $k \ll
\pi$ in lattice units, are rotationally invariant and very close to the
correct ``hard thermal loops'', except for corrections which are
$O(\omega^2)$ and $O(Q^4 \beta_{\rm L}^2 \langle n \rangle)$ (both in
lattice units).  This is
totally unlike the hard thermal loops induced by hard lattice modes in
the nonabelian theory or the abelian Higgs model, which are
rotationally non-invariant \cite{Smilga}.

\section{Diffusion of Chern-Simons number}
\label{NCSsec}

An outstanding question which the method developed here can answer is;
what is the diffusion constant for Chern-Simons number $N_{\rm CS}$ in the
symmetric electroweak phase, or in pure Yang-Mills theory?  In
the continuum, the Chern-Simons number is defined as
\begin{eqnarray}
\partial^\mu K_\mu & \equiv & \frac{g^2}{32 \pi^2} \frac{\epsilon^{\mu \nu
	\alpha \beta}}{2} F_{\mu \nu}^a F_{\alpha \beta}^a \, , \\
N_{\rm CS} \equiv \int d^3 x K_0 & = & \frac{g^2}{32 \pi^2} \int d^3 x
	\epsilon_{ijk} \left( F^a_{ij} A^a_k - \frac{g}{3} f_{abc}
	A^a_i A^b_j A^c_k \right) \, .
\end{eqnarray}
where Latin indices $(ijk)$ imply sums over the three space directions with
positive metric.  In vacuum $N_{\rm CS}$ is an integer, equal to the
winding number of the gauge transformation which carries the gauge
fields into the trivial fields $A=0$.  This can be nonzero for groups
with nontrivial third homotopy group.  Chern-Simons number is of
interest physically because of how it is related to the behavior of
fermions coupled to the gauge fields; a vacuum to vacuum process which
changes $N_{\rm CS}$ by $n$ pulls $n$ left handed negative energy
solutions of the Dirac operator up to positive energy and pushes $n$
right handed solutions from positive to negative energy, if the
fermion couples to the gauge fields in the fundamental
representation.  In a chiral theory, like the SU(2) sector of the
standard model, no right handed particles couple, and there will be
net particle creation (recall that in vacuum, negative energy
solutions are occupied and positive energy solutions are not).
Summing over the $N_{\rm F}=3$ generations one finds that the baryon 
number and lepton number currents both have nonzero divergences,
\begin{equation}
\partial^\mu J_{\mu {\rm L}} = \partial^\mu J_{\mu {\rm B}} 
	= N_{\rm F} \partial^\mu K_\mu \, , 
\end{equation}
so the number of baryons $N_{\rm B}$ and of leptons $N_{\rm L}$ changes as
\begin{equation}
N_{\rm B}(t) - N_{\rm B}(0) 
	= N_{\rm L}(t) - N_{\rm L}(0) 
	= N_{\rm F} ( N_{\rm CS}(t) - N_{\rm CS}(0) ) \, .
\end{equation}
At high temperatures the efficiency with which $N_{\rm B}$ is violated is
related to the diffusion constant for $N_{\rm CS}$, 
\begin{equation}
\Gamma \equiv \lim_{V \rightarrow \infty} \lim_{t \rightarrow \infty}
	\frac{ \langle ( N_{\rm CS}(t) - N_{\rm CS}(0))^2 \rangle}{Vt}
\end{equation}
(where $\langle \rangle$ refers to the thermal ensemble) by a 
fluctuation-dissipation relation \cite{KhlebShap,RubakShap2} 
and standard thermodynamic arguments \cite{ArnoldMcLerran}.  In the
minimal standard model\footnote{In extensions with light 
baryon number or lepton number carrying particles, $39/4$
will be replaced by something smaller; in the supersymmetric theory, if
all the squarks and sleptons were light, $39/4$ would 
become $39/12$.}, one finds
\begin{equation}
\frac{1}{N_{\rm B} + N_{\rm L}} \frac{d(N_{\rm B} + N_{\rm L})}{dt} 
	= \frac{39}{4T^3} \Gamma \, .
\end{equation}
It would be phenomenologically interesting to know $\Gamma$
at high temperatures, which presumably existed in the early universe
before the electroweak SU(2)$\times$U(1) symmetry was spontaneously
broken.

There has been a great deal of work to date on determining $\Gamma$
\cite{AmbKras,MooreTurok,TangSmit,Ambjornetal,Moore1},
but recently there have been two important developments.  

The first is an analytic argument due to Arnold, Son, and Yaffe (ASY)
\cite{ArnoldYaffe}.  They point out that a change in $N_{\rm CS}$ involves
the evolution of very infrared magnetic fields.  The hard thermal loops
cause these fields to evolve in an overdamped manner.  This is
familiar from the study of abelian (electromagnetic) plasmas; magnetic
fields of wavelength $\lambda \gg 1/\omega_{\rm p}$ get ``frozen'' by the
conductivity of the plasma and evolve on the time scale $\tau \sim
\lambda (\lambda \omega_{\rm p})^2$, assuming $\lambda$ is much shorter
than the diffusion length of the charge carriers.  ASY 
argue that the same physics should apply in the nonabelian
plasma, at least for $\lambda$ less than or on order $1/g^2 T$; so
$\Gamma$ should be parametrically of order 
\begin{equation}
\Gamma^{-1} \sim \lambda^3 \tau \sim \left( \frac{1}{g^2T} \right)^3
	\frac{\omega_{\rm p}^2 }{ (g^2 T)^3} \, .  
\end{equation}
If this argument is correct then $\Gamma$ depends strongly on the
physics of hard thermal loops, which we must get right to find the
correct $\Gamma$.  Classical lattice theory by itself (without classical
particles) does not.  The damping coefficient describing the 
overdamped evolution of
infrared magnetic fields on long time scales grows linearly 
with $1/a$ and is not rotationally invariant \cite{Arnoldlatt}.

The second development is that better definitions of $N_{\rm CS}$ on the
lattice have been developed \cite{slavepaper,AmbKras2,slave3}.  
Previous definitions in terms of local operators contained
lattice spacing dependent systematic errors. Moore and Turok 
\cite{slavepaper} proposed a definition which is topological and
hence avoids such errors. Their results verify that $\Gamma$ depends
strongly on lattice spacing, in a manner which appears consistent with
the ASY scaling law, though only if there are substantial
corrections to that law at higher order in $ (g^2 T)^2 / \omega_{\rm p}^2 $.

By using the topological method of Moore and Turok to track the
evolution of $N_{\rm CS}$, and by using the classical field theory with
particles added to correctly reproduce the hard thermal loops, we can
now get a determination of the diffusion constant for $N_{\rm CS}$ which
accounts correctly both for topology and for hard thermal loops.  We can
also ensure that the thermodynamics of the system under study are
correct by using the $O(a)$ improved matching developed in
\cite{Oapaper} and extended to arbitrary Debye mass in Appendix
\ref{AppendixB}.  

We will do so in pure Yang-Mills theory, which should correspond to the
very high temperature limit of the Standard Model because the thermal
Higgs mass becomes large enough at high temperature that the Higgs field
can be integrated out (though its contribution to the hard thermal loops
should of course be included).  The sphaleron rate in the symmetric
phase at the equilibrium point of the phase transition will differ
somewhat from the Yang-Mills theory value, in a 
way which depends on the (unknown) couplings of the Higgs
sector.  It is straightforward to add the Higgs field to the theory we
have developed--in particular the influence of hard modes on the Higgs
field can be completely accounted for by the choice of Higgs mass
\cite{Aarts}--so there is no obstacle to extending what we do here to
that case.

There is an important complication to our plan; in a nonabelian
theory, the UV classical field modes will also generate hard thermal 
loops; and as B{\"o}deker et. al. \cite{Smilga} have shown, the 
functional form of the hard thermal loops they provide is NOT the same 
as Eq. (\ref{HTL}).  Hence, the actual hard thermal loop
contribution to the lattice system with particles will be of form
\begin{equation}
m^2_{\rm {D, particles}} \cdot \Gamma_{\rm particles}[A] + 
m^2_{\rm {D, UV \, lattice}} \cdot \Gamma_{\rm UV \, lattice}[A] \, ,
\end{equation}
where $\Gamma_{\rm particles}[A]$ has the correct and 
$\Gamma_{\rm UV \, lattice}[A]$ has the wrong functional form.
This is further complicated because of the interactions between 
the particles and the UV lattice modes.  The UV lattice modes are Landau
damped, and as we increase $m_{\rm D}^2$ from particles, that damping becomes
stronger.  If this damping is strong enough, then UV classical modes
have short propagation distances, and they will not propagate 
interactions over large spatiotemporal separations.  The
hard thermal loop effects most important to the $N_{\rm CS}$ diffusion 
rate are those on length scales of order $1/g^2 T$
\cite{ArnoldYaffe}, so the relevant contribution from UV classical 
lattice modes may have an $m_{\rm D}^2$ 
dependent suppression.  We do not know
a good way to estimate the importance of this suppression, so we will
treat it as a source of systematic error.

We want to know the result of a triple limit.  The innermost
limit is the limit of $Q^2 \rightarrow 0$ and $\langle n \rangle
\rightarrow \infty$ with $Q^2 \langle n \rangle$ fixed; in this limit
the particles generate only hard thermal loop effects.  The next limit
is the limit as $m^2_{\rm {D, particles}} / m^2_{\rm {D, UV \, lattice}}$
becomes large; in this limit the hard thermal loops are of the
correct functional form, plus a correction which is small relative to
the total strength.  At fixed lattice spacing, this limit
means making $m_{\rm D}^2$ large; so we can only learn about the
parametrically leading behavior in the limit of large $m_{\rm D}^2$.
Finally, we should take a small lattice spacing limit.  

Limited numerical resources make it impossible to really achieve
this triple limit.  We will choose a value for $Q^2$ small enough that
we can expect to be in the relevant limit there, and we will also
assume that $\beta_L \sim 10$ puts us far enough in the small $a$
limit, if $O(a)$ thermodynamic corrections are used\footnote{We verify in
Appendix \protect{\ref{AppendixA}} that the frequency corrections to the
particle hard thermal loops are $O(\omega^2 a^2)$, and we believe the
same should be true of the finite $k$ corrections.  This leaves
nonrenormalizable operators, also $O(k^2 a^2)$, thermodynamic errors
beyond $O(a)$, and an $O(a)$ rescaling of the time scale, which we
estimate in Appendix \protect{\ref{AppendixD}} but have not computed.}.  
We will check both of these limits by varying $Q^2$ holding $Q^2 \langle
n \rangle$ fixed and by varying $a$, to verify that the dependence is
weak; but we make no serious attempt to extrapolate to these limits.  We
concentrate on what we consider the most phenomenologically interesting
limit, what happens as we make $m_{\rm D}^2$ large.  In particular we want to
know whether the $\Gamma_d$ scales according to the ASY prediction, 
$\Gamma_d \propto m_{\rm D}^{-2}$.

We will then try to check three things:
\begin{enumerate}
\item 
	$\Gamma$, in physical units, should depend weakly on the lattice
	spacing, provided that the physical value of the plasma
	frequency is held fixed;
\item
	$\Gamma$ should depend on 
	$Q^2$ and $\langle n \rangle$ only through the combination $Q^2
	\langle n \rangle$, when we have chosen $Q^2$ small enough that the
	particle trajectories are ballistic on the nonperturbative
	scale $(g^2T)^{-1}$;
\item
	$\Gamma$ should depend inversely on hard thermal loop strength,
	up to corrections due to the UV classical lattice mode 
	contributions.
\end{enumerate}

\begin{figure}[t]
{\centerline{\mbox{\psfig{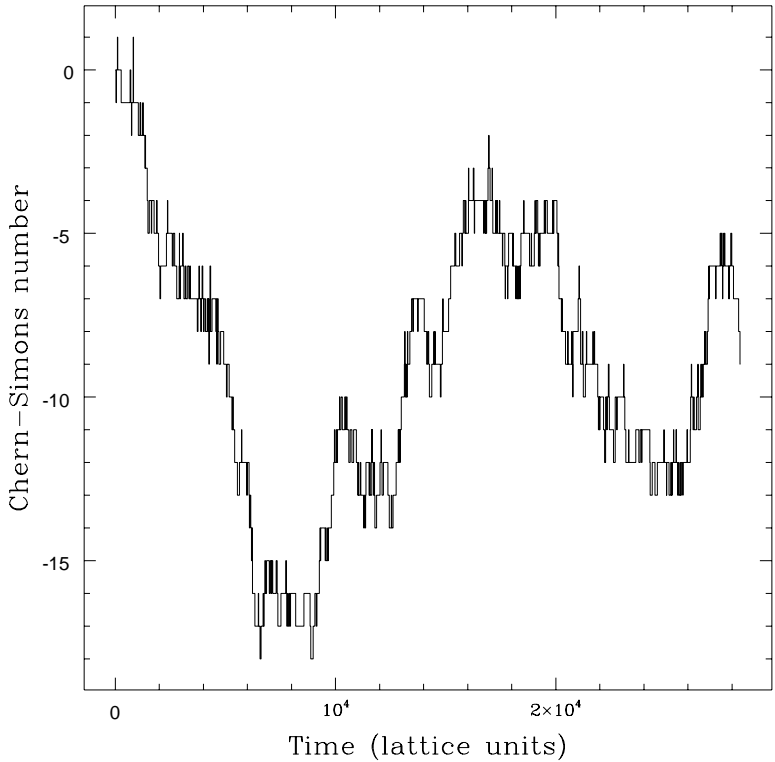}}}}
\caption{ \label{NCSsample}
$N_{\rm CS}$ during a Hamiltonian evolution, tracked by the slave field
method (which returns the integer value of $N_{\rm CS}$ for the nearest
vacuum).  This is part of the data for row (5) of the table.
The back and forth stuttering accompanying many winding number
changes is an expected feature if the ASY picture is right
\protect{\cite{ArnoldYaffe}}. }
\end{figure}

To test $1.$, we measure $\Gamma$ at three lattice spacings, but with the
same value of $Q^2$ and $\langle n \rangle$ in physical units.  To test
$2.$, we double the value of $\langle n \rangle$ and halve the value of
$Q^2$ relative to the runs used in $1.$, and to test $3.$ we 
double and quadruple $\langle n \rangle$ relative to 
the runs used to study $1.$, but keep the same
value of $Q^2$.  If the ASY scaling law is right, then up to corrections
due to HTL's from UV classical lattice modes, all
these runs should give the same value for $\Gamma$, except for the last,
where $\Gamma$ should be about half 
and then a quarter as large.  If $\Gamma$ does not depend
on hard thermal loops, then all the results should be the same, and
should be the same as the result without any particles added.

\begin{table}
\centerline{\mbox{\begin{tabular}{|c|c|c|c|c|c|} \hline
$\beta_{\rm L, imp}$ (improved) & lattice size & $\langle n \rangle$ 
& $Q^2$ & $\kappa$ & $\kappa'$ \\ \hline
8  & $20^3$  & 58.6 & .0064  &   $.84 \pm .08$ & $49 \pm 5 \pm 7$ \\ \hline
10 & $24^3$  & 30.0 & .0064  &   $.92 \pm .07$ & $56 \pm 4 \pm 10$ \\ \hline
12 & $30^3$  & 17.4 & .0064  &   $.72 \pm .07$ & $46 \pm 5 \pm 10$ \\ \hline
10 & $24^3$  & 60.0 & .0032  &   $.72 \pm .06$ & $44 \pm 4 \pm 8$ \\ \hline
10 & $24^3$  & 60.0 & .0064  & $.475 \pm .041$ & $53 \pm 5 \pm 5$ \\ \hline
10 & $24^3$  &120.0 & .0064  & $.249 \pm .025$ & $53 \pm 5 \pm 2$ \\ \hline
\end{tabular}}}
\caption{\label{NCStable}
$N_{\rm CS}$ diffusion constant in physical units, $\Gamma = \kappa 
\alpha^4 T^4$, varying lattice spacing, particle charge, 
and particle number.  The inverse lattice spacing 
$\beta_{\rm L} = 4 / g^2 a T$ used here is the one
including the perturbative corrections found in \protect{\cite{Oapaper}}
and Appendix \protect{\ref{AppendixB}}.
The last column is the coefficient of the ASY scaling law, see the text.}
\end{table}

Our results are presented in Table \ref{NCStable}.  Each data point is
extracted from several Hamiltonian evolutions from independent thermal
initial conditions.  We present a sample of $N_{\rm CS}$ during such a
Hamiltonian trajectory in Figure \ref{NCSsample}.
The analysis techniques used to extract $\Gamma$ are the same as in
\cite{slavepaper}.  For each choice of parameters the sum of lengths of
evolutions is about 90000 lattice units of time.
The value of $\langle n \rangle$ for the $\beta_{\rm L} = 8, 10, 12$
data keeps  the physical particle density, which is proportional to 
$\beta_{\rm L}^3 \langle n \rangle$, fixed; and as we have discussed, $Q^2$ 
does not scale with lattice spacing.  Hence these three results test
(1).  Row 4 has the same value of $Q^2 \langle n \rangle$ as row 2, so
comparing them tests (2).  Finally, comparing rows 2, 5, and 6 tests (3).

The results are expressed through the dimensionless quantity $\kappa$,
which in the continuum is defined through 
\begin{equation}
\Gamma = \kappa \alpha^4 T^4 \, ,
\end{equation}
or on the lattice,
\begin{equation}
\Gamma = \kappa (\beta_{\rm L, imp} \pi)^{-4}
\end{equation}
times a correction, discussed in Appendix \ref{AppendixD}, to account
for the correct matching of the time scales.

We also present the results in terms of the coefficient of the ASY
scaling law, which we write as
\begin{equation}
\Gamma = \kappa' \left( \frac{g^2 T^2}{m_{\rm D}^2} \right) \alpha^5 T^4 \, .
\end{equation}
We use $m_{\rm D}^2$ here because it most conveniently 
characterizes the size of hard thermal loop effects.  
For the particle degrees of freedom, the Debye mass in physical units is
\begin{equation}
m_{\rm D}^2 = \frac{Q^2 \langle n 
	\rangle({\rm latt. \; units}) \beta_{\rm L}^3 
	g^4 T^2}{48} \, , 
\label{zztops}
\end{equation}
which for the first 4 columns equals $4 g^4 T^2$, a little less than the
physical value, which is $11 g^2 T^2 / 6$, $g^2 \simeq 0.4$ \cite{KLRS}.

Really the ASY scaling law says the results should depend not on the
Debye mass but on a damping coefficient proportional to the transverse
self-energy at $\omega \ll k \sim g^2 T$.  This is simply related 
to $m_{\rm D}^2$ in the case that the hard particles have a
rotationally invariant spectrum and move at the speed of light.
The particle degrees of freedom satisfy this requirement, but the hard
classical lattice modes do not, so we have accounted for their
contribution using the techniques of Arnold \cite{Arnoldlatt}.  His
result is that the ratio of damping coefficient to Debye mass 
squared is roughly $(0.68 \pm 0.2)$ times smaller for hard classical 
lattice modes than for ultrarelativistic particles.  However,
Landau damping of the UV classical lattice modes, mentioned earlier, may
suppress their contribution to the transverse self-energy at 
$k \sim g^2 T$; since we do not know how to compute the extent of this
suppression, there is a systematic error.  The upper limit of the
systematic error bar we present is if they contribute 
fully, in which case we add
$0.68$ times the Debye mass squared from classical lattice
modes to that from particles when converting $\kappa$ to $\kappa'$.
(In physical units the Debye mass squared from classical lattice modes
is $m_{\rm D}^2({\rm latt}) = ( \Sigma
\beta_{\rm L} / 4 \pi) g^4 T^2$.)   The lower limit is
if their contribution is fully frustrated by Landau damping off of hard
particles, in which case we just use $m_{\rm D}^2$ from particles to convert
from $\kappa$ to $\kappa'$.  The larger $Q^2 \langle n \rangle$, the 
stronger the Landau damping; so the systematic is not
common to all runs.

Our results are roughly consistent with lattice spacing independence 
and with dependence on $Q^2$ and $\langle n \rangle$ only through 
the combination $Q^2 \langle n \rangle$.
There does seem to be a weak systematic trend in lattice spacing, which
could be partly from $O(a^2)$ effects we have
not attempted to compute.  For instance, besides the $O(a)$ corrections
to the thermodynamics computed in \cite{Oapaper} and Appendix
\ref{AppendixB}, there are two $O(a^2)$ corrections: a renormalization
of the coupling, and a nonrenormalizable $(D_i F_{ij})^2$ term, which
appears in the Hamiltonian with a negative sign.  Both would raise the
rate on coarser lattices.  There are also $O(a^2)$ corrections to the
Wong's equation limit of the interactions between particles and long
wavelength modes.  There also may be a weak trend in $Q^2$ when $Q^2
\langle n \rangle$ is held constant, because we are not sufficiently 
close to the small $Q^2$ and large $\langle n \rangle$ limit.  
Not being in this limit means that the
eikonal approximation used to turn the particles into hard thermal loop
effects is not quite true.  Scattering of the particles will tend to
reduce their effectiveness.  Hence one might expect a weak systematic
where $\kappa$ rises with $Q^2$ at $Q^2 \langle n \rangle$ fixed.  Row 4
in the table suggests this but the effect is not very statistically
significant.  Systematics have not been eliminated but they are small.

\begin{figure}[t]
{\centerline{\mbox{\psfig{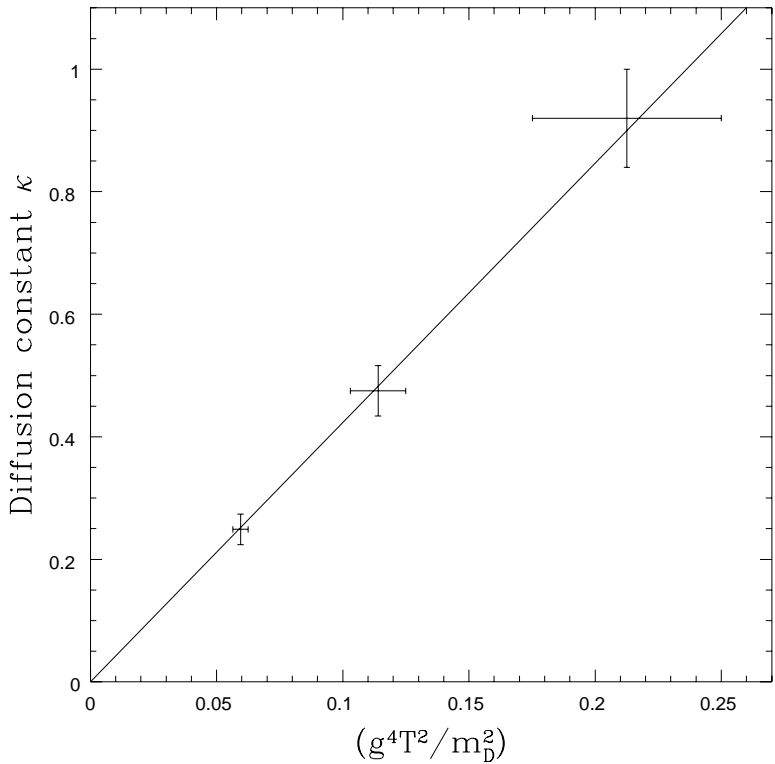}}}}
\caption{ \label{theyagree}
Results for $\kappa$ at three values of particle number but a common
value of lattice spacing, plotted against $g^4 T^2/m_{\rm D}^2$.  
The error bars in $m_{\rm D}^2$ reflect 
uncertainty in the damping from hard classical lattice
modes.  The ``old picture'' predicts a flat line, while the ASY picture
predicts a straight line through the origin, like the illustrated fit.}
\end{figure}

Our results rule out the ``old picture'' that $\Gamma$ should depend 
on $\alpha^4 T^4$ by demonstrating 
the importance of hard thermal loop effects.  
If the $\alpha^4 T^4$ law were correct, all the values for $\kappa$
would agree, and would agree with the value in Yang-Mills theory without
particles, which is $\kappa = 1.53 \pm 0.10$ at $\beta_{\rm L} = 10$
\cite{slavepaper}.  The three results, rows (2), (5), and (6) of the
table, differ only in the number of particles used, not in the lattice
spacing, the particle charge, or the manner in which finite $\beta_{\rm
L}$ lattice spacing systematics were taken care of.  They are grossly in
conflict if $\kappa$ is independent of $m_{\rm D}^2$, but they agree very
nicely with the ASY scaling law.  We illustrate this in Figure
\ref{theyagree}.  Combining them, we get an estimate for $\kappa'$ of
$\kappa' = 53 \pm 3_{\rm stat} \pm 5_{\rm syst}$.  We should also fold
in the systematic errors from finite lattice spacing and finite $Q^2$,
which we would estimate based on the other runs to be in the $20\%$
range.  These then dominate our uncertainties, and our final answer is
$\kappa' = 53 \pm 11$.
For comparison, we can take the results for pure Yang-Mills
theory without particles from \cite{slavepaper} and extrapolate them
to zero lattice
spacing assuming the ASY scaling applies.  The extrapolation is $\kappa
= 23.6 / \beta_{\rm L}$, see figure 7 of that paper.  We can convert
this into a value for $\kappa'$ by using the discussion after
Eq. (\ref{zztops}), that is, using Arnold's calculation of the relation
between damping from hard classical lattice modes and from correct hard
thermal loops \cite{Arnoldlatt}.  We get
$\kappa' = 51 \pm 15 $.\footnote{We thank Peter Arnold and Jan Smit for
discussions on this point.} 
The error here is almost all systematic, arising from
the rotational noninvariance of the spectrum of lattice modes.

It is encouraging that these results agree within (admittedly
substantial and mainly systematic) errors.
It appears that a consistent picture for the $N_{\rm CS}$ diffusion
constant has emerged.

\section{Conclusions}
\label{conclusionsec}

We have developed a procedure for generating the nonlocal ``hard
thermal loop'' effects in classical field simulations of Yang-Mills
theory, by introducing particle degrees of freedom which generate
these effects but can be treated with reasonable numerical effort.
The new system is Hamiltonian and has the same thermodynamics as
thermal Yang-Mills theory in the dimensional reduction approximation.
We have also tested that the self-energy corrections due to
the particles in the abelian theory are correct, and we have 
analytically computed the plasma frequency at leading order and
it is also correct.

Further, we have applied the technique to the calculation of the
diffusion constant of Chern-Simons number in pure Yang-Mills theory.
Our results vindicate the proposed scaling law of Arnold, Son, and
Yaffe, that the diffusion constant $\Gamma$ should scale 
inversely with the square of the Debye mass.  This statement is not
sensitive to the quality of the thermodynamic corrections we have
applied, because we have data with different numbers of particles 
which show starkly different diffusion rates, but the lattice spacing
and hence the thermodynamic corrections we have applied are the same.
Our results (in Yang-Mills theory, which will be valid 
in Yang-Mills Higgs theory only for 
thermal Higgs mass $m_H^2(T) \gg g^4 T^2$) convert to 
$\Gamma = 29 \pm 6 \alpha^5 T^4$ at $m_{\rm D}^2 = 11 g^2 T^2 / 6$.

There are no obstacles to extending our techniques to Yang-Mills
Higgs theory or groups larger than SU(2), which would allow us
to determine the screening dependence of a number of other dynamical
properties of interest for baryogenesis, such as the IR bosonic
contribution to the bubble wall friction and the strong sphaleron rate.
The technique may also
have applications for relativistic heavy ion collisions.

\section*{Acknowledgments}
We would like to acknowledge Peter Arnold, Dam Son,
Misha Shaposhnikov, Keijo Kajantie, Jan Smit, Dirk Rischke, 
and Sergei Matinyan for useful conversations.  GM is 
grateful to Duke University for hospitality during a brief visit there.
This work was supported in part by the U.S. Department of Energy (Grant
No. DE-FG02-96ER40495) and by the North Carolina Supercomputing Center
and the Pittsburgh Computing Center.

\appendix

\section{Plasma frequency in the abelian theory}
\label{AppendixA}

We have shown that the thermodynamic modifications due to particles
are very close to those expected, and we have argued that the role of
the particles in generating hard thermal loops should be the same as
in the continuum theory, for suitably infrared and slowly varying
gauge fields.  Here we will check this, and test its limits, by
explicitly calculating the plasma frequency, the oscillation frequency
of a spatially homogeneous electric field, in the abelian theory.  The
calculation should apply approximately to the nonabelian theory, in
the regime where $\omega_{\rm p} \gg g^2 T$, since in this case the
theory looks approximately abelian on the relevant length scales.
Away from this limit, the plasma frequency of the nonabelian theory is
not well defined, since the electric fields only oscillate coherently
on a time scale of order $1/g^2 T$.  We will calculate in the formal
small $Q$, large $\langle n \rangle$, but finite $Q^2 \langle n \rangle$
limit, but we will deal explicitly with the discrete nature of the
lattice electric fields and their interactions with particles.

First we will consider the case of a continuous electric field.  It is
sufficient to consider an electric field pointing along a lattice
direction, since we will work to linear order in the field strength,
so general fields can be studied as a linear combination.  We
take the field to be $E_i(t) = \delta_{ix} 
{\rm Re} E_0 e^{i \omega t}$, with $E_0$ a constant giving the
strength of the electric field, and we
will solve self-consistently for $\omega$.  Consider a particle
of charge $q \ll 1$
propagating in the background of this field, with a mean momentum in
the $x$ direction of $p_x$ and a momentum orthogonal to the $x$
direction of $p_{\perp}$.  The $x$ momentum will have a time dependent
disturbance of $\delta p_x(t) \propto q E_0$, 
due to the electric field, which satisfies 
\begin{equation}
\delta \dot{p}_x = q E_x = {\rm Re} \, q E_0 e^{i \omega t} 
	\quad \rightarrow \quad
	\delta p_x = {\rm Re} \frac{ - i q E_0}{\omega} e^{i \omega
	t} \, .
\end{equation}
The electric field responds to a current in the $x$ direction due to
this particle, of magnitude $q v_x$.  Expanding in $\delta p_x \ll 1$,
This is
\begin{equation}
v_x = \frac{p_x + \delta p_x}{\sqrt{p_\perp^2 + p_x^2 + 2 p_x \delta p_x}}
	\simeq \frac{p_x}{p} + \frac{\delta p_x}{p} \left( 1 - \frac{p_x^2}
	{p^2} \right) = \frac{p_x}{p} + 
	\frac{\delta p_x p_\perp^2}{p^3} \, .
\end{equation}
Now averaging over particles, the leading term will cancel between
particles of opposite sign for $q$, and the time derivative of the
electric field will be
\begin{equation}
- \dot{E}_x = \frac{1}{V} \sum_{\rm particles} 
	\frac{q \delta p_x p_\perp^2}{p^3} = {\rm Re} \frac{-i
	q^2 E_0}{\omega} e^{i \omega t} \sum_{\rm particles}
	\frac{p_\perp^2}{p^3} \, .
\end{equation}
Now $\dot{E}_x = {\rm Re} i \omega E_0 e^{i \omega t}$.  The $E_0$
cancel on the two sides, and we get an expression for $\omega^2$.  A
nice way of thinking of this is that what we want to know is ${\rm Re}
( - i \dot{E} / E ) = \omega$; we just plug the determined value of
$\dot{E}$ into this expression to find $\omega$.

Now we evaluate the expression.
The average over angles of $p_\perp^2$ is $2p^2/3$, and the average
over particles of $1/p$ is $1/2T$.
\begin{equation}
\omega^2 = \frac{ \langle n \rangle \langle q^2 \rangle }{3T} \, .
\end{equation}
In lattice units, $1/T$ becomes $\beta_{\rm L}$.  In the nonabelian case,
$\langle q^2 \rangle = Q^2/3$, because that is the average of the
square of the projection of a particle charge into one abelian
subgroup.  We therefore find $\omega_{\rm p}^2 
= m_{\rm D}^2/3$, as it should.

Now let us include the discrete nature of the electric
fields and their interactions with the particles.  Again we take the
electric field to be spatially uniform with the same value, and we
will solve self-consistently for the plasma frequency.  

Again, what we want to know is ${\rm Re} ( - i \dot{E} / E )$, or its time
average.  We consider the influence of one particle, then we will
average over all particles.  Again the particle has a mean momentum in
the $x$ direction of $p_x$ and a perpendicular momentum of $p_\perp$.
We will often write $\pm$ to mean ${\rm sign}(p_x)$ and $\mp$ to mean
$- {\rm sign}(p_x)$.

The impulse on the electric field on some link, due to a particle
crossing the dual lattice face the link penetrates, is of fixed
modulus $\mp q$.  
Past interactions with $E$ change the
current the particle induces only by changing
the time at which the impulse is felt.\footnote{There will also be a
small number of particles which would have induced a current on this
link, but instead induce a current on its neighbor.  But there are an
equal number which induce a current on this link rather than its
neighbor, and this cancels out.}
Let the particle passage time, at zero order in $q$, be $t_{\rm
cross}$.  This crossing time will receive a correction due to past
interactions of the particle with the electric field, which we denote
$\delta t(t_{\rm cross})$.  The particle's contribution to 
$ {\rm Re} \int dt (-i\dot{E}/E)$ is 
\begin{eqnarray}
E_0 {\rm Re} \int dt (-i\dot{E}/E) & 
	= & {\rm Re} \left[ (\pm i q) e^{-i \omega
	(t_{\rm cross} + \delta t(t_{\rm cross}))} \right] \\
	& = & {\rm Re} \left[ (\pm i q) e^{-i \omega t_{\rm cross}} \right]
	+ {\rm Re} \left[ \pm q \omega \delta t(t_{\rm cross}) 
	e^{-i \omega t_{\rm cross}} \right] \, . \nonumber
\end{eqnarray}
The leading term is $q$ odd, and cancels between particles of opposite
charge.  But $\delta t(t_{\rm cross})$ will be $q$ odd at leading order,
and this term will contribute at $O(q^2)$.

Let us compute $\delta t(t)$ to leading order in $q$.  
It is related to the
difference in the particle's position from the $E=0$ case, $\delta
x(t)$, by $\delta t(t) = - \delta x(t) |p|/p_x$.  This in turn is given
by an integral over the particle's past history of the difference in
its velocity from the average value,
\begin{equation}
\delta x(t) = \int_{- \infty}^t dt' \delta v_x(t') = 
 	\int_{- \infty}^t dt' \delta p_x(t') \frac{p_{\perp}^2}{p^3} \, .
\end{equation}
Now $\delta p_x(t')$ is a sum over past crossings of the kick to the
particle at that crossing.  The particle has undergone an infinite
number of past crossings, one for each plane parallel the plane of the
dual lattice face it is currently crossing.  The time since it crossed
the plane a distance $\mp m$ away in the $x$ direction is $\pm m |p| /
p_x$, and the phase of the electric field at that time was $\exp(i
\omega (t_{\rm cross} \mp m |p| / p_x))$, and the impulse it received
was, at leading order in $q$, $\pm q E |p|/p_x$.  Hence,
\begin{equation}
\delta x(t_{\rm cross})) = {\rm Re} \left( e^{i \omega t_{\rm cross}}
	\sum_{m=1}^{\infty} \frac{m|p|}{\pm p_x} \frac{q E_0 |p|}
	{\pm p_x} \frac{p_{\perp}^2}{p^3} 
	e^{ \mp i \omega m |p|/p_x}  \right) \, .
\end{equation}

This sum is not absolutely
convergent.  So to make sense of the calculation, we must regulate
it.  We do this by assuming that, as well as fluctuating with the
plasma frequency, the correlator of $E$ at two times has a slow
exponential envelope, so $E_x$ in the past is not ${\rm Re} E_0 e^{i
\omega t}$ but ${\rm Re} E_0 e^{(i \omega + \epsilon) t}$.  Of course,
we will take the $\epsilon \rightarrow 0$ limit at the end.

Thus, after regulating the sum, the kick
from this particle crossing this boundary is
\begin{equation}
{\rm Re} \int dt \frac{-i \dot{E}}{E} = {\rm Re} \sum_{m=1}^{\infty}
	\mp \frac{m q^2 \omega p_\perp^2}{p_x^3} e^{\mp (i \omega +
	\epsilon) m |p| / p_x} \, .
\end{equation}
The frequency with which a particle of this momentum encounters a
boundary in the $x$ direction is $\pm p_x / p$, so $\omega$ is
\begin{equation}
\omega = \frac{1}{V} \sum_{\rm particles} {\rm Re} \sum_{m=1}^{\infty}
	- \frac{m q^2 \omega p_\perp^2}{p_x^2 |p|} e^{\mp (i \omega +
	\epsilon) m |p| / p_x} \, .
\end{equation}
Now writing $x = p_x/|p|$, and performing the integral over $p^2 dp$
and over the aximuthal angle, this reduces to
\begin{equation}
\omega^2 = \frac{ \langle n \rangle \langle q^2 \rangle \beta_{\rm L}}
	{3} {\rm Re} \lim_{\epsilon \rightarrow 0} \frac{3 \omega^2}{2}
	\int_0^1 dx \frac{x^2-1}{x^2} \sum_{m=1}^{\infty} m e^{(-i
	\omega - \epsilon) m/x} \, .
\end{equation}
The sum over $m$ may now be performed,
\begin{equation}
\sum_{m=1}^{\infty} m e^{(i \omega - \epsilon) m/x} = 
	\frac{1}{4} \sinh^{-2} \left( \frac{ \epsilon + i \omega }
	{2 x} \right) \, .
\end{equation}

Making the substitution $y=1/x$, the integral becomes
\begin{equation}
\omega^2 = \frac{ \langle n \rangle \langle q^2 \rangle \beta_{\rm L}}
	{3} {\rm Re} \lim_{\epsilon \rightarrow 0} \frac{3 \omega^2}{2}
	\int_1^\infty dy \frac{1-y^2}{4 y^2} \sinh^{-2} 
	\left( \frac{y ( \epsilon + i \omega )}{2} \right) \, .
\end{equation}
For $\epsilon > 0$ the integral is exponentially convergent, and we
can rotate the contour to run in the negative imaginary direction:
\begin{equation}
\omega^2 = \frac{ \langle n \rangle \langle q^2 \rangle \beta_{\rm L}}
	{3} {\rm Re} \lim_{\epsilon \rightarrow 0} \frac{3 \omega^2}{2}
	\int_0^\infty i dy \frac{ (y+i)^2 + 1}{4 (y+i)^2} 
	\sinh^{-2} \left( \frac{ (y+i)(\omega - i \epsilon)}{2}
	\right) \, .
\end{equation}
The integral is well behaved and we are free to take the $\epsilon
\rightarrow 0$ limit.  Taking the real part, after some algebra we
obtain 
\begin{eqnarray}
\omega^2 & = & \frac{ \langle n \rangle \langle q^2 \rangle \beta_{\rm L}}
	{3} \times F(\omega^2) \, , \\
F(\omega^2) & = & \frac{3 \omega^2}{2} \int_0^\infty dy
	\frac{2y (\cosh(\omega y) \cos(\omega) - 1) + (y^4 + 3y^2)
	\sinh(\omega y) \sin(\omega)}{2 (1+y^2)^2 (\cosh(\omega y) - 
	\cos(\omega) )^2} \\
	& \simeq & 1 - \frac{\omega^2}{4}
	+ \frac{\omega^4}{240} + \frac{\omega^6}{30240} \ldots \,
	. \nonumber 
\end{eqnarray}
Here, of course, $\omega$ is the plasma frequency in lattice units, ie
$a \omega$ in physical units.
We see that it receives a correction when $\omega \sim 1/a$, but that
for $\omega^2 a^2 \sim 1/2$ in lattice units, the correction is already
quite small.  There will also be $O(\langle n \rangle Q^4)$
corrections, probably including damping, but we have not been able to
calculate these analytically.

\section{Integrating out the $A_0$ field at arbitrary $m_{\rm D}$}
\label{AppendixB}

We would like to understand the relation between the thermodynamics of
infrared magnetic fields in our lattice system and the thermodynamics
of infrared magnetic fields in the full quantum theory, say in the
dimensional reduction approximation.  As we have seen, our system has
thermodynamics described by a lattice gauge theory with an $A_0$
field, which has a Debye mass $m_{\rm DL}^2$ which depends on $\beta_{\rm L}$
and on $\langle n \rangle Q^2$, whereas the dimensional reduction
approximation gives the continuum limit of this lattice theory, with
some particular renormalized $m_{\rm D}^2$ determined by the particle
content of the full quantum field theory.  There are two complications
here, the difference in behavior between lattice and continuum systems
and the difference in Debye mass.  The first has been dealt with in
\cite{Oapaper}, for the system without the $A_0$ field and for the
system with the $A_0$ field but in the approximation that $m_{\rm DL}^2$
is small in lattice units.  In the theory with only classical lattice
modes this approximation is parametrically justified, but with the
inclusion of particles this is not necessarily the case.

To deal with finite $m_{\rm D}^2$, one first notes that $m_{\rm D}$ is large
enough to make the $A_0$ field heavy, and since the theory is
super-renormalizable one can integrate over such heavy fields at one
loop and capture their dominant contributions to the infrared.  We
should perform this integration in each system, leaving us to compare
pure Yang-Mills theory in 3-D, on the lattice and in the continuum.
The matching between these has been studied in \cite{Oapaper} up to
corrections of order $\beta_{\rm L}^{-2}$, so here we only discuss the
integration over the $A_0$ field.

Integrating over the $A_0$ field in the continuum theory was studied
in \cite{KLRS,FKRS}.  There is only one correction, a self-energy
correction to the gauge fields which shifts the gauge coupling $g_3^2
\equiv g^2 T$ to 
\begin{equation}
\bar{g}_3^2 = g_3^2 \left( 1 - \frac{g_3^2}{24 \pi m_{\rm D}} \right) \, .
\label{contingcorr}
\end{equation}
In the theory with a Higgs field there  is also a shift to the scalar
self-coupling, 
\begin{equation}
\bar{\lambda}_3 = \lambda_3 - \frac{3 g_3^4}{128 \pi m_{\rm D}} \, ,
\label{lambdacorr}
\end{equation}
and to the Higgs mass squared parameter, 
\begin{equation}
\bar{m}_3^2  = m_3^2 - \frac{3 g_3^2 m_{\rm D}}{16 \pi} + O(g_3^4 / 16
	\pi^2) \, .
\end{equation}
The scalar wave function is not corrected.

In the lattice theory the integrals are trickier because the $A_0$
field has lattice dispersion relations and the gauge-$A_0$ vertices
have nontrivial momentum dependence.  So to warm up we will start with the
corrections to the Higgs parameters.  Our unit conventions will be the
same as in \cite{Oapaper}, ie the lattice scalar self-coupling will be
$\lambda_{\rm L} = 4 \lambda / g^2$, and the lattice spacing will appear
in the dimensionless quantity $\beta_{\rm L} = 4 / ( g^2 a T)$.

\begin{figure}[t]
\centerline{\psfig{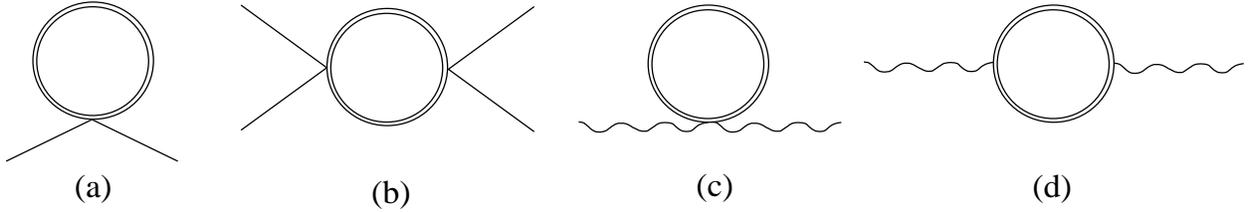}}
\caption{\label{A0diagrams}
Diagrams needed in the integration over the $A_0$ field in Yang-Mills
Higgs theory.  The double solid line is the $A_0$ field, the single
solid line is the Higgs field, and the wavy line is the gauge field.}
\end{figure}

Two diagrams are relevant, Fig. \ref{A0diagrams} diagrams $(a)$ and
$(b)$.  There are also high dimension operators induced by diagrams
with more $\phi$ lines, but the influence of these terms is small for
large $m_{\rm D}^2$ and we can drop them, just as in the continuum case
\cite{KLRS}.  Diagram $(a)$ gives a momentum independent self-energy
correction, ie a mass squared correction.  Denote the coupling
between the $A_0$ field and the Higgs field as $\lambda_{\rm A}$;
at lowest order $\lambda_{\rm AL} = 2$.  Here and in what follows an $L$
subscript means the value in lattice units, with the normalizations
used in \cite{Oapaper}.  The shift in the lattice
Higgs mass is\footnote{Note that there are several
typos in the Appendix C of \cite{Oapaper} in which factors of $\Sigma
/ 4 \pi$ or $\xi / 4 \pi$ are left out.}
\begin{equation}
\bar{m}_{\rm HL}^2 = m_{\rm HL}^2 + \frac{3 h_{\rm AL}}{2 \beta_{\rm L}}
	\int_{[-\pi , \pi]^3} \frac{d^3 k}{(2 \pi)^3}
	\frac{1}{\tilde{k}^2 + m_{\rm DL}^2} \, ,
\label{eq91}
\end{equation}
where $\tilde{k}_i \equiv 2 \sin( k_i / 2)$ and 
\begin{equation}
\tilde{k}^2 = \sum_i \tilde{k}_i^2  = \sum_i ( 2 - 2 \cos k_i ) \, .
\end{equation}
Henceforth the range of integration and the $d^3 k/(2
\pi)^3$ will be understood.

The integral in (\ref{eq91}) must be determined numerically, and we 
define it as 
\begin{equation}
\int_k \frac{1}{\tilde{k}^2 + m_{\rm DL}^2} \equiv \frac{\Sigma(m_{\rm DL})}
	{4\pi} \longrightarrow
	\frac{\Sigma}{4 \pi} - \frac{m_{\rm DL}}{4 \pi} - \frac{\xi
	m_{\rm DL}^2}{4 \pi} + \frac{m_{\rm DL}^3}{32 \pi} + \ldots \, ,
\end{equation}
where $\Sigma = \Sigma(0) = 3.175911536$ and $\xi = 0.152859325$.  We
display the small $m_{\rm DL}$ expansion, but at general $m_{\rm DL}$ the
integral should be done numerically.

At zero external momentum diagram $(b)$ corrects the scalar
self-coupling by
\begin{equation}
\bar{\lambda}_{\rm L} = \lambda_{\rm L} 
	- \frac{3 h_{\rm AL}^2}{4 \beta_{\rm L}} \int_k
	\frac{1}{( \tilde{k}^2 + m_{\rm DL}^2)^2} \, .
\end{equation}
We define
\begin{equation}
\int_k \frac{1}{( \tilde{k}^2 + m_{\rm DL}^2)^2} 
	\equiv \frac{ \xi(m_{\rm DL}) }
	{4\pi} \longrightarrow \frac{1}{8 \pi
	m_{\rm DL}} + \frac{\xi}{4 \pi} - \frac{3 m_{\rm DL}}{64 \pi} 
	+ \ldots \, .
\end{equation}
Again, we have displayed the leading terms at small
$m_{\rm DL}$, but at general $m_{\rm DL}$ the integral should be treated
numerically. 
This definition of $\xi(m_{\rm DL})$ 
does not have $\xi(m \rightarrow 0) \rightarrow \xi$;
instead $\xi(m)$ has a pole at $m=0$, and $\xi$ is the constant part
after the pole.  The residue of the pole reproduces the continuum
expression, (\ref{lambdacorr}), and the presence of the pole is
because we are computing the full effect of the $A_0$ field on the
lattice, not the difference between lattice and continuum theories.

Now we treat gauge field self-energy corrections.  Since
there are no $A_0$ field corrections to the three point gauge-scalar
vertex or to the Higgs wave function
at one loop, this is the only further correction we need.
There are two diagrams, $(c)$ and $(d)$ of Figure \ref{A0diagrams}.
Denoting the polarization and spin indicies as $i,j$ and $a,b$, 
at general external momentum $p$, diagram $c$ gives
\begin{equation}
- \frac{8 \delta_{ab} \delta_{ij}}{\beta_{\rm L}} \int \frac{ \cos k_i}
	{\tilde{k}^2 + m_{\rm DL}^2} \, , 
\label{diagramc}
\end{equation}
and diagram $(d)$ gives
\begin{equation}
\frac{\delta_{ab}}{\beta_{\rm L}} \int \frac{ 16 \sin k_i \sin k_j} 
	{ [ ( \widetilde{k-p/2} )^2 + m_{\rm DL}^2 ] 
	[ ( \widetilde{k+p/2} )^2 + m_{\rm DL}^2 ] } \, .
\label{diagramd}
\end{equation}
The sum of the contributions at $p=0$ is
\begin{equation}
- \frac{\delta_{ab}}{\beta_{\rm L}} \int \left( \frac{8 \delta_{ij} \cos k_i}
	{\tilde{k}^2 + m_{\rm DL}^2} - \frac{ 16 \sin k_i \sin k_j } 
	{ ( \tilde{k}^2 + m_{\rm DL}^2 )^2} \right)
	= - 4 \frac{\delta_{ab}}{\beta_{\rm L}}
	\int \frac{\partial^2}{\partial k_i \partial k_j}
	\ln ( \tilde{k}^2 + m_{\rm DL}^2 ) = 0 \, .
\label{identity1}
\end{equation}
The last equality is because the integrand is a total derivative
without singularities and the domain of integration is compact without
boundary.  Hence the $A_0$ field does not induce a mass for the gauge
field, as ensured by gauge invariance.

Next, we find the $O(p^2)$ term in the self-energy, which is
responsible for renormalizing the coupling.  Only diagram $(d)$
depends on $p$, and expanding (\ref{diagramd}) to 
second order gives
\begin{equation}
\frac{ 4 \delta_{AB}}{\beta_{\rm L}} \sum_{lm} p_l p_m \left( 
	-\int \frac{2 \sin k_i \sin k_j \cos k_l \delta_{lm}}
	{(\tilde{k}^2 + m_{\rm DL}^2)^3}
	+ \int \frac{4 \sin k_i \sin k_j \sin k_l \sin k_m}
	{(\tilde{k}^2 + m_{\rm DL}^2)^4} \right) \, .
\end{equation}
To evaluate this, we need the identity
\begin{eqnarray}
0 & = & \int \frac{ \partial^2}{\partial k_i \partial k_j} 
	\frac{\cos k_l}{\tilde{k}^2 + m_{\rm DL}^2}  \nonumber \\
	& = & \int \frac{8 \sin k_i \sin k_j \cos k_l}
	{(\tilde{k}^2 + m_{\rm DL}^2)^3}
	- \frac{2 \delta_{ij} \cos k_i \cos k_l}
	{(\tilde{k}^2 + m_{\rm DL}^2)^2}
	+ \frac{ \cos k_i \delta_{ijl}}
	{\tilde{k}^2 + m_{\rm DL}^2} \, ,
\label{identity2}
\end{eqnarray}
where we have already used (\ref{identity1}) to simplify the
final expression, and the identity
\begin{eqnarray}
0 & = & \int \frac{ \partial^4 }{\partial k_i \partial k_j \partial k_l
	\partial k_m} \ln ( \tilde{k}^2 + m_{\rm DL}^2) \nonumber \\
	& = & - \int \frac{96 \sin k_i \sin k_j \sin k_l \sin k_m}
	{(\tilde{k}^2 + m_{\rm DL}^2)^4} + \int 16 \frac{ \delta_{ij} \cos k_i
	\sin k_l \sin k_m + 5 \: {\rm permutations} }{(\tilde{k}^2 + 
	m_{\rm DL}^2)^3} + \nonumber \\
	& & + \int \frac{6 \delta_{ijkl} \cos k_i}{\tilde{k}^2 + m_{\rm DL}^2}
	- \int 4 \frac{\cos k_i \cos k_l \delta_{ij} \delta_{lm} + 2 \:
	{\rm permutations}}{(\tilde{k}^2 + m_{\rm DL}^2)^2} \, . 
\label{identity3}
\end{eqnarray}
Further, we use the relationship
\begin{equation}
\sum_i \cos k_i = \left( 3 + \frac{m_{\rm DL}^2}{2} \right) - 
	\frac{1}{2} \left( \tilde{k}^2 +  m_{\rm DL}^2 \right) \, ,
\end{equation}
from which it follows that
\begin{equation}
\int \frac{ \cos k_i }{\tilde{k}^2 +  m_{\rm DL}^2} = - \frac{1}{6} + 
	\left( 1 + \frac{m^2}{6} \right) \frac{\Sigma(m_{\rm DL})}{4 \pi}
	\, ,
\label{identity4}
\end{equation}
and, with a little more work, that
\begin{eqnarray}
\! \! \int \frac{\cos k_i \cos k_j}{(\tilde{k}^2 +  m_{\rm DL}^2)^2}
	& \! = \! & \left[ \left( 1 + \frac{m_{\rm DL}^2}{2} 
	+ \frac{m_{\rm DL}^4}{24} \right)
	\frac{\xi(m_{\rm DL})}{4 \pi} - \left( \frac{1}{4} 
	+ \frac{m_{\rm DL}^2}{24} \right)
	\frac{\Sigma(m_{\rm DL})}{4 \pi} \right] + \nonumber \\
	& & + \delta_{ij} \left[ - \left( \frac{m_{\rm DL}^2}{2} 
	+ \frac{m_{\rm DL}^4}{24} \right) \frac{\xi(m_{\rm DL})}{4 \pi} -
	\left( \frac{1}{4} + \frac{m_{\rm DL}^2}{24} \right)
	\frac{\Sigma(m_{\rm DL})}{4 \pi} + \frac{1}{12} \right] .
\label{identity5}
\end{eqnarray}
Using these, after considerable algebra we find that the $O(p^2)$
contribution to the self-energy is
\begin{equation}
\frac{\delta_{ab}}{\beta_{\rm L}} \left( \delta_{ij} p^2 - p_i p_j \right)
	\left[ \left( - \frac{4}{3} - \frac{2m_{\rm DL}^2}{3} - 
	\frac{m_{\rm DL}^4}{18} \right) \frac{ \xi(m_{\rm DL})}{4 \pi} +
	\left( \frac{1}{3} + \frac{m_{\rm DL}^2}{18} \right) \frac{
	\Sigma(m_{\rm DL})} {4 \pi} \right] \, .
\label{mainresult}
\end{equation}
While intermediate expressions have been rotationally non-invariant,
the result is rotationally invariant and transverse.  This happens
because the only gauge invariant dimension 4 operator which is cubic
invariant is $F_{ij}^2$.  At $O(p^4)$ there will be rotationally
non-invariant contributions, although they will be cubic invariant.
We are not concerned with these here because we are not trying to
compute induced nonrenormalizable operators, only corrections to terms
already in the Hamiltonian.  But the nonrenormalizable terms 
could be computed by a
straightforward but extremely tedious extension of what we have done
here. 

The $m_{\rm DL} = 0$ limit of (\ref{mainresult}) agrees with the result in
\cite{Oapaper}, and the coefficient of the pole at $m_{\rm DL} = 0$
reproduces (\ref{contingcorr}).  In the opposite limit of 
very large $m_{\rm DL}$ one can (Taylor) expand $\Sigma$ and $\xi$ in
$m_{\rm DL}^{-2}$, 
\begin{eqnarray}
\frac{\Sigma(m)}{4 \pi} & \rightarrow & m^{-2} - 6 m^{-4} + 42 m^{-6}
	- 324 m^{-8} + 2610 m^{-10} - \ldots \, , \\
\frac{\xi(m)}{4 \pi} & \rightarrow & m^{-4} - 12 m^{-6} + 126 m^{-8}
	- 1296 m^{-10} + 13050 m^{-12} - \ldots \, .
\end{eqnarray}
The expansion converges for $m^2 > 12 = {\rm sup} (\tilde{k}^2)$.
While the $\xi$ and $\Sigma$ contributions in (\ref{mainresult})
separately give $O(m^0)$ contributions, these cancel, as do the first
several powers of $m^2$, and the correction is $8 \delta_{ab} (p^2
\delta_{ij} - p_i p_j) / ( \beta_{\rm L} m_{\rm DL}^8)$ at leading order in
$m_{\rm DL}^{-2}$.  So the $A_0$ field rapidly becomes irrelevant at large
Debye mass, as it should.

Combining the result from integrating
out the $A_0$ field with the thermodynamic corrections from transverse
gauge boson loops found in \cite{Oapaper}, we get a relation between
the ``naive'' tree level $\beta_{\rm L,naive}$ and the improved value,
\begin{equation}
\beta_{\rm L,naive} = \beta_{\rm L,imp} + \left( \frac{1}{3} + \frac{37
	\xi}{6 \pi} \right) + 	\left[ \left( - \frac{4}{3} - 
	\frac{2m_{\rm DL}^2}{3} - 
	\frac{m_{\rm DL}^4}{18} \right) \frac{ \xi(m_{\rm DL})}{4 \pi} +
	\left( \frac{1}{3} + \frac{m_{\rm DL}^2}{18} \right) \frac{
	\Sigma(m_{\rm DL})} {4 \pi} \right] \, .
\end{equation}
Here the constant $\xi = 0.152859325$ is the limit of $\xi(m)$ as $m$
goes to zero, after the pole has been removed.

\begin{table}
\centerline{\mbox{\begin{tabular}{|c|c|c|c|c|c|} \hline
$\beta_{\rm L,naive}$ & $Q^2 \langle n \rangle$ & $m_{\rm D}^2$ &
$\Sigma(m_{\rm D})$ & $\xi(m_{\rm D})$ & $\beta_{\rm L,imp}$ \\ \hline
8.7  & .375  & 1.59 & 1.901 & .3487 & 8.073  \\ \hline
10.7 & .189  & 1.09 & 2.101 & .4556 & 10.078 \\ \hline
12.7 & .111  & 0.74 & 2.280 & .5821 & 12.085 \\ \hline
10.7 & .378  & 1.77 & 1.840 & .3214 & 10.072 \\ \hline
10.7 & .756  & 3.14 & 1.498 & .1979 & 10.069 \\ \hline
\end{tabular}}}
\caption{\label{appBtable}
Conversion between $\beta_{\rm L,naive}$ and $\beta_{\rm L,imp}$ for the
simulations in this paper.  Everything is in lattice units.
In Section \protect{\ref{NCSsec}} we
truncate $\beta_{\rm L,imp}$ to the nearest integer when we write it.}
\end{table}

For completeness, we also list the values of $\beta_{\rm L,imp}$ and 
$\beta_{\rm L, naive}$ for the simulations presented in section 
\ref{NCSsec}, in Table \ref{appBtable}.

This completes the integration over the $A_0$ field at one loop, at
general $m_{\rm DL}^2$.

\section{Probing the retarded propagator by linear response}
\label{AppendixC}

In this appendix we show how to probe the retarded photon propagator by 
studying the linear response of the abelain plasma to an external current. 

First we consider the response to transverse perturbations.  
In the framework of linear response theory, the response $A(x)$
is related to the external current $j(x)$ through the retarded propagator
$D^R(x,x^{\prime})$:
\begin{equation}
A_i(x)=\int d^4 x^{\prime}D_{ik}^R(x,x^{\prime}) j_k(x^{\prime})~.
\end{equation}
Going to Fourier space and suppressing the Lorentz indices, we have
\begin{equation}
\tilde{E}(\omega,k)=i\omega\tilde{D}^R(\omega,k)\tilde{j}(\omega,k)~,
\end{equation}
where
\begin{equation}
\tilde{D}^R(\omega,k)=\frac{-1}{\omega^2-k^2-\Pi(\omega,k)} \, .
\end{equation}
Taking $j_i(\vec{x},t)=\hat{j}_i j_0 e^{-i\omega_0 t} 
e^{ik \cdot x}\theta(t)$ as the driving current, we find
\begin{equation}
\tilde{E}_i(\omega,l)= \hat{j}_i \frac{ 
	j_0 \omega (2\pi)^3 \delta^3(\vec{k} - \vec{l})}
	{(\omega-\omega_0+i\epsilon)(\omega^2-k^2-\Pi(\omega,k))} \, .
\end{equation}
Fourier transforming $\tilde{E}(\omega,k)$ back to space-time gives
\begin{equation}
E(\vec{x},t) = j_0 e^{ik_0 x}\int\frac{d\omega}{2\pi}
	\frac{\omega e^{-i\omega t}}
	{(\omega-\omega_0+i\epsilon)(\omega^2-k^2-\Pi(\omega,k))} \, .
\end{equation}
The retarded propagator $D^R(\omega,k)$ is analytic in the upper
half complex $\omega$-plane due to causality, and only has poles 
in the lower half plane. Writing 
$\Pi(\omega,k)=\Pi_{\rm r}(\omega,k)+i\Pi_{\rm i}(\omega,k)$
and noting that 
$\Pi_{\rm r}(-\omega,k)=\Pi_{\rm r}(\omega,k)$ 
and $\Pi_{\rm i}(-\omega,k)=-\Pi_{\rm i}(\omega,k)$, 
the two poles are located at
$\omega_{\rm pl}(k_0) = \omega_{\rm p}(k_0) -i\gamma_{\rm p}(k_0)$ and
$-\omega^*_{\rm pl}(k_0)$,
where $\omega_{\rm p}(k_0)$ is the plasma frequency for the mode $k_0$
and $\gamma_{\rm p}(k_0)$ is the associated on-shell damping rate.

Completing the contour integral in the lower half $\omega$-plane, we find
\begin{eqnarray}
E(\vec{x},t) &=&  -i \frac{ j_0\omega_0 e^{-i\omega_0 t} e^{ik_0 x} }
	{\omega_0^2-k_0^2-\Pi(\omega_0,k_0)} \theta(t)  \nonumber \\
& &  	-i \frac{ j_0 e^{-\gamma_{\rm p}(k_0)t} e^{ik_0 x} } 
	{ \omega_{\rm pl}(k_0)+\omega_{\rm pl}^*(k_0) }
	\left[ \frac{ \omega_{\rm pl}(k_0) e^{-i\omega_{\rm p}(k_0)t} } 
	{\omega_{\rm pl}(k_0)-\omega_0}
	-\frac{ \omega_{\rm pl}^*(k_0) e^{i\omega_{\rm p}(k_0)t} } 
	{\omega_{\rm pl}^*(k_0)+\omega_0}
	\right] \theta(t)~.
\end{eqnarray}
The first term represents the asymptotic response, while the second
term is the plasma oscillation at $k=k_0$, which is only transient due to
damping at a rate of $\gamma_{\rm p}(k_0)$. 
At large $t$ ($t\gg \gamma_{\rm p}^{-1}(k_0)$), after the transients have
died out, one sees only the asymptotic behavior
\begin{equation}
E(\vec{x},t) \longrightarrow 
	-i \frac{ j_0\omega_0 e^{-i\omega_0 t} e^{ik_0 x} }
	{\omega_0^2-k_0^2-\Pi(\omega_0,k_0)}~.
\end{equation}

A current of the form 
$j(\vec{x},t)=j_0\sin{(\omega_0 t)}\sin{(k_0 x)}\cdot\theta(t)$
will generate the following $E$ field at large $t$,
\begin{equation}
E(\vec{x},t) = j_0\omega_0\rho
	\cos{(\omega_0 t+\alpha)}\sin{(k_0 x)}\cdot\theta(t)~,
\end{equation}
where
\begin{eqnarray}
\rho &=& \left\{[\omega_0^2-k_0^2-\Pi_{\rm r}(\omega_0,k_0)]^2 
	+ \Pi_{\rm i}^2(\omega_0,k_0)\right\}^{-\frac{1}{2}}~,    \\
\alpha &=& \tan^{-1}{\left[\frac{-\Pi_{\rm i}(\omega_0,k_0)}
	{\omega_0^2-k_0^2-\Pi_{\rm r}(\omega_0,k_0)}\right]}~.
\end{eqnarray}
The real and imaginary part of $\Pi(\omega,k)$ can therefore be
determined from the space-time averages
\begin{equation}
\langle j(\vec{x},t)\cdot E(\vec{x},t) \rangle
	=\frac{j_0^2}{4}\cdot \frac{\omega_0 \Pi_{\rm i}(\omega_0,k_0)}
	{[\omega_0^2-k_0^2-\Pi_{\rm r}(\omega_0,k_0)]^2 
	+ \Pi_{\rm i}^2(\omega_0,k_0)}~,
\label{part1}
\end{equation}
and
\begin{equation}
\langle j(\vec{x},t)\cdot E(\vec{x},t-\tau) \rangle
	=\frac{j_0^2}{4}\cdot
	\frac{\omega_0 [\omega_0^2-k_0^2-\Pi_{\rm r}(\omega_0,k_0)]}
	{[\omega_0^2-k_0^2-\Pi_{\rm r}(\omega_0,k_0)]^2 
	+ \Pi_{\rm i}^2(\omega_0,k_0)}~,
\label{part2}
\end{equation}
where $\tau=\pi/(2\omega_0)$, and the averages are taken after the
decay of the transients.

The numerically determined values for the transverse self-energies
$\Pi_{\rm i}$ and $\Pi_{\rm r}$ can then
be compared to their perturbative values \cite{KlimovWeldon}:
\begin{eqnarray}
\Pi_{\rm r}(\omega,k) &=& \frac{3}{2}\omega_{\rm p}^2\frac{\omega^2}{k^2}
	\left[ 1 + \frac{1}{2}(\frac{k}{\omega}-\frac{\omega}{k})
	\ln{\left|\frac{\omega+k}{\omega-k}\right|}\right]~,  \\
\Pi_{\rm i}(\omega,k) &=& - \frac{3 \pi}{4}
	\omega_{\rm p}^2\frac{\omega^2}{k^2}
	(\frac{k}{\omega}-\frac{\omega}{k})\theta(k^2-\omega^2)~.
\end{eqnarray}

The response to longitudinal perturbations is easiest to treat in Coulomb
gauge, where only $A_0$ is nonzero and $E$ is determined from $\partial_i
A_0$.  The retarded propagator is
\begin{equation}
D^{\rm R}_{00} ( \omega , k ) = \frac{- ( \omega^2 - k^2)}
	{k^2 (\omega^2 - k^2 - \Pi_{\rm L} ( k , \omega ))} \, .
\end{equation}
One obtains
\begin{eqnarray}
\langle j ( x , t ) \cdot E ( x , t ) \rangle & = & 
	\frac{j_0^2}{4 \omega} 
	\frac{(\omega^2 - k^2) \Pi_{\rm L,i}}
	{ ( \omega^2 - k^2 - \Pi_{\rm L,r} )^2 + \Pi_{\rm L,i}^2 } \, , \\
\langle j ( x , t ) \cdot E ( x , t - \tau ) \rangle & = & 
	\frac{j_0^2}{4 \omega}
	\frac{(\omega^2 - k^2) (\omega^2 - k^2 - \Pi_{\rm L,r} ) }
	{ (\omega^2 - k^2 - \Pi_{\rm L,r} )^2 + \Pi_{\rm L,i}^2 } \, .
\end{eqnarray}
The real and imaginary parts of the longitudinal self-energy are
\begin{eqnarray}
\Pi_{\rm L,r}(\omega , k) & = & 3 \omega_{\rm p}^2 (1-\frac{\omega^2}{k^2})
	\left(
	1 - \frac{ \omega}{2k} \ln \left| \frac{ \omega + k }
	{\omega - k} \right| \right) \, , \\
\Pi_{\rm L,i} (\omega , k) & = & \frac{3 \pi}{2} \omega_{\rm p}^2 
	\frac{\omega}{k} (1-\frac{\omega^2}{k^2})
	\theta ( k^2 - \omega^2 ) \, .
\end{eqnarray}

\section{Relating time scales when there are particles}
\label{AppendixD}

It has been shown \cite{Oapaper}, as discussed in Appendix
\ref{AppendixB}, that the difference in screening from UV modes between
lattice gauge theory and continuum gauge theory can be understood at
leading order in $a$ (or in $\beta_{\rm L}^{-1}$) as a rescaling of the
differential operator $D_i$, so what at tree level looks like 
$D^2 A$ in fact behaves like $Z_A^{-1} D^2 A$, with $Z_A = 1 + O(a)$.
The $O(a)$ term has been computed in \cite{Oapaper} and extended
to arbitrary Debye mass in Appendix \ref{AppendixB}.  Hence, in the case
with no particles, the equations of motion for $A_i$ in temporal gauge
look like
\begin{eqnarray}
\frac{\partial E_i}{\partial t} 
& = & - Z_A^{-1} ( D^2 \delta_{ij} - \frac{1}{2} ( D_i D_j 
	+ D_j D_i ) ) A_j \, , \label{Eevolution} \\
\frac{\partial A_i}{\partial t} & = & Z_E E_i \, , 
\end{eqnarray}
and $\langle E^2 \rangle \propto \beta^{-1}_{\rm L, naive}$.  The
usual rule $\dot{A} = E$ is rescaled by $Z_E$, which has not been
computed, but which lacks the large ``tadpole'' contributions which
characterize $Z_A$.  To get a complete $O(a)$ correction of the dynamics
it would be necessary to compute $Z_E$, but it has been advocated in
\cite{slavepaper} that the absence of ``tadpole'' contributions means
that $Z_E - 1$ can be neglected compared to $Z_A - 1$.  In this case, a
simple rescaling of time, $Z^{1/2}_A d/dt = d/dt'$, and of the
electric field, $Z^{1/2}_A E = E'$, eliminates $Z_A$, and replaces
$\beta_{\rm L, naive}$ with $\beta_{\rm L, imp} = Z_A^{-1}
\beta_{\rm L, naive}$.  So the time scale used, $t$, is related to the
more appropriate time scale, $t'$, by $t' = t \sqrt{ \beta_{\rm L, imp}
/ \beta_{\rm L, naive}}$.  When one computes $\Gamma$ one should
divide $\langle N_{\rm CS}^2 \rangle$ by $V t'$ rather than $Vt$, and the
corrected rate is  $\sqrt{ \beta_{\rm L, naive}
/ \beta_{\rm L, imp}}$ times larger.

However we are now interested in the case where there are enough
particles to put the evolution in the overdamped regime.  Huet and Son 
\cite{HuetSon} and Son \cite{Son} have recently shown what 
new term this adds to ( \ref{Eevolution} ).  A thermal distribution
of particles would not contribute, since particles of opposite charge
would cancel, but the particle distribution at any point is skewed
because of past, remote electric fields, and contains fluctuations.
The new evolution equation looks like
\begin{eqnarray}
\frac{\partial E_i^a(x,t)}{\partial t} + Z_A^{-1} ( D^2 \delta_{ij} 
	- \frac{1}{2} ( D_i D_j + D_j D_i ) ) A_j^a (x,t) 
	= 
\frac{Q^2 \langle n \rangle \beta_{\rm L, naive}}{3} 
   \nonumber \\
\times
\int dy N_{ij}^{ab}(x,y,t) E^b_j (y,t-|x-y|) + ({\rm noise \; term}) \, .
\end{eqnarray}
Here the right hand side represents the departure of the particle
population from rotational invariance, due to its linear response to
past electric fields, and fluctuations in the particle population.  The
form of the nonlocal kernel $N$ is given in \cite{HuetSon,Son}.  They
point out that the overdamped limit corresponds to the large $Q^2
\langle n \rangle$ limit, in which case the ${\partial E}/{\partial t}$ 
term and the time
dependence of $E$ on the right hand side can be neglected.  Again
approximating that $\dot{A} = E$, rescaling time to eliminate $Z_A$ and
to replace all appearances of $\beta_{\rm L, naive}$ with
$\beta_{\rm L, imp}$ now requires $t' = Z_A^{-2} t$.  Hence one should
compute everything using $\beta_{\rm L, imp}$ and the lattice spacetime
volume used, but then correct the rate by a factor of $(\beta_{\rm L,
naive} / \beta_{\rm L, imp})^2$.



\begin{thebibliography}{99}

\bibitem{Sakharov} A. Sakharov, JETP Lett. {\bf 6}, 24 (1967).
\bibitem{tHooft} G. t'Hooft, Phys. Rev. Lett. {\bf 37},8 (1976).
\bibitem{Manton} F. Klinkhamer and N. Manton, Phys. Rev. {\bf D 30},
         2212 (1984).
\bibitem{KRS85}V. Kuzmin, V. Rubakov, and M. Shaposhnikov, Phys. 
        Lett. {\bf D 30}, 36 (1985).
\bibitem{ArnoldMcLerran} P. Arnold and L. McLerran, Phys. Rev. {\bf D 36},
        581 (1987).
\bibitem{oldDR} P. Ginsparg, Nucl. Phys. {\bf B 170}, 388 (1980);
	T. Applequist and R. Pisarski, Phys. Rev. {\bf D 23}, 2305 (1981);
	S. Nadkarni, Phys. Rev. {\bf D 27}, 917 (1983);
	N. P. Landsmann, Nucl. Phys. {\bf B 322}, 498 (1989).
\bibitem{FKRS1} K. Farakos, K. Kajantie, K. Rummukainen, 
        and M. Shaposhnikov, Nucl. Phys. {\bf B 425}, 67
        (1994).
\bibitem{KLRS} K. Kajantie, M. Laine, K. Rummukainen, and M.
	Shaposhnikov, Nucl. Phys. {\bf B 458}, 90 (1996).
\bibitem{FKRS} K. Farakos, K. Kajantie, K. Rummukainen, 
        and M. Shaposhnikov, Nucl. Phys. {\bf B 442}, 317
        (1995).
\bibitem{Laine} M. Laine, Nucl. Phys. {\bf B 451}, 484 (1995).
\bibitem{Oapaper} G. D. Moore, Nucl. Phys. {\bf B 493}, 439 (1997).
\bibitem{KLRSresults}
	K. Kajantie, M. Laine, K. Rummukainen, and M. Shaposhnikov,
	Nucl. Phys. {\bf B 466}, 189 (1996);
	Phys. Rev. Lett. {\bf 77}, 2887 (1996).
\bibitem{Kripfganz} H. Dosch, J. Kripfganz, A. Laser, and M. Schmidt,
	Phys. Lett. {\bf B 365}, 213 (1996);
	M. Gurtler, E. Ilgenfritz, J. Kripfganz, H.
	Perlt, and A. Schiller, Nucl. Phys. {\bf B 483}, 383 (1997);
	M. Gurtler, E. Ilgenfritz, and A. Schiller, Eur. Phys. J. 
	{\bf C 1}, 363 (1998); Phys. Rev. {\bf D 56}, 3888 (1997).
\bibitem{Teper} O. Philipsen, M. Teper, and H. Wittig, Nucl. Phys.
	{\bf B 469}, 445 (1996); A. Hart, O. Philipsen, J. Stack, and M.
	Teper, Phys. Lett. {\bf B 396}, 217 (1997).
\bibitem{MooreTurok} G. D. Moore and N. Turok, Phys. Rev. {\bf D 55}, 
	 6538 (1997).
\bibitem{KLRSSU2U1} K. Kajantie, M. Laine, K. Rummukainen, and M.
	Shaposhnikov, Nucl. Phys. {\bf B 493}, 413 (1997).
\bibitem{AmbKras} J. Ambj{\o}rn and A. Krasnitz, Phys. Lett. {\bf B 362},
         97 (1995).
\bibitem{GrigRub} D. Grigorev and V. Rubakov, Nucl. Phys. {\bf B 299},
         248 (1988).
\bibitem{Bodeker} D. B{\"o}deker, Nucl. Phys. {\bf B 486}, 500 (1997).
\bibitem{Smilga} D. B{\"o}deker, L. McLerran, and A. Smilga, Phys.
        Rev. {\bf D 52}, 4675 (1995).
\bibitem{KhlebShap} S. Khlebnikov and M. Shaposhnikov, Nucl. Phys.
        {\bf B308}, 885 (1988).
\bibitem{TangSmit} W. Tang and J. Smit, Nucl. Phys. {\bf B 482}, 265 (1996).
\bibitem{ArnoldYaffe} P. Arnold, D. Son, and L. Yaffe, Phys. Rev.
	{\bf D 55}, 6264 (1997).
\bibitem{HuetSon} P. Huet and D. Son, Phys. Lett. {\bf B393}, 94 (1997).
\bibitem{slavepaper} G. D. Moore and N. Turok, Phys. Rev. {\bf D 56},
	 6533 (1997).
\bibitem{AmbKras2} J. Ambj{\o}rn and A. Krasnitz, Nucl. Phys. 
	{\bf B 506}, 387 (1997).
\bibitem{Geiger} K. Geiger, Phys. Rep. {\bf 258}, 237 (1995).
\bibitem{Wang} X. N. Wang, Phys. Rep. {\bf 280}, 287 (1997).
\bibitem{Thoma} M. H. Thoma, in Quark-Gluon Plasma 2, edited by R. C. Hwa,
	(World Scientific, Singapore, 1995), P. 51.  
\bibitem{HuMuller} C. Hu and B. M{\"u}ller,
 	Phys. Lett. {\bf B409}, 377 (1997).
\bibitem{KellyMIT94} P.F. Kelly, Q. Liu, C. Lucchesi, and C. Manuel, 
        Phys. Rev. Lett. {\bf 72}, 3461 (1994);
        Phys. Rev. {\bf D 50}, 4209 (1994).
\bibitem{BraatenPnucl90} E. Braaten and R.D. Pisarski, 
        Nucl. Phys. {\bf B 337}, 569 (1990).
\bibitem{BraatenPprd90} E. Braaten and R.D. Pisarski, 
	Phys. Rev. {\bf D 42}, 2156 (1990).
\bibitem{TaylorWong90} J.C. Taylor and S.M.H. Wong, 
        Nucl. Phys. {\bf B 346}, 115 (1990).
\bibitem{BlaizotIprl93} J.P. Blaizot and E. Iancu, 
        Phys. Rev. Lett. {\bf 70}, 3376 (1993).
\bibitem{HeinzElze89} H. Elze and U. Heinz,
        Phys. Rep. {\bf 183}, 81 (1989).
\bibitem{Wong70}
        S.K. Wong, Nuo. Cim. {\bf A 65}, 689 (1970).
\bibitem{Heinz84} U. Heinz, Phys. Lett. {\bf B 144}, 228 (1984).
\bibitem{EfratyNair92} R. Efraty and V.P. Nair, 
        Phys. Rev. Lett. {\bf 68}, 2891 (1992). 
\bibitem{Kogut} J. Kogut and L. Susskind, 
	Phys. Rev. {\bf D 11}, 395 (1975).
\bibitem{Ambjornetal} J. Ambj{\o}rn, T. Askgaard, H. Porter, and M.
             Shaposhnikov, Nucl. Phys. {\bf B 353}, 346 (1991).
\bibitem{Moore1} G. D. Moore,  Nucl. Phys. {\bf B 480}, 657 (1996);
	Nucl. Phys. {\bf B 480}, 689 (1996).
\bibitem{Krasnitz} A. Krasnitz, Nucl. Phys. {\bf B 455}, 320 (1995).
\bibitem{RubakShap2}  V. Rubakov and M. Shaposhnikov,
	Phys. Usp. {\bf 39}, 461 (1996)
	[Usp. Fiz. Nauk {\bf 166}, 493 (1996)].
\bibitem{Arnoldlatt} P. Arnold, Phys. Rev. {\bf D 55}, 7781 (1997).
\bibitem{slave3} G. D. Moore, Phys. Lett. {\bf B 412}, 359 (1997).
\bibitem{Aarts} G. Aarts and J. Smit, Phys.  Lett. 
	{\bf B 393}, 395 (1997).
\bibitem{KlimovWeldon} 
        V. Klimov, Sov. J. Nucl. Phys. {\bf 33}, 934 (1981);
	A. Weldon, Phys. Rev. {\bf D 26}, 1394 (1982). 
\bibitem{Son} D. Son, UW/PT-97-19, hep-ph/9707351.
\end{thebibliography}
\end{document}